\documentclass[iop,numberedappendix]{emulateapj}
\slugcomment{Accepted by \apj} 
\usepackage{natbib}
\usepackage{amsmath}
\usepackage{booktabs}
\usepackage{longtable}
\usepackage{color}
\usepackage{graphicx, subfigure}
\usepackage{hyperref}
\newcommand{\di}{\mathrm{d}} 
\newcommand{\mr}{\mathrm} 
\newcommand{\Ht}{\mathrm{H_2}} 
\newcommand{\CO}{\mathrm{CO}}

\newcommand{\pc}{\mathrm{pc}} 
\newcommand{\Myr}{\mathrm{Myr}}

\defcitealias{KO2017}{KO2017}

\begin{document}
\title{The $X_\CO$ conversion factor from galactic multiphase ISM simulations}
\author{Munan Gong\altaffilmark{1,2}, Eve C. Ostriker\altaffilmark{1}
and Chang-Goo Kim\altaffilmark{1,3}}
\altaffiltext{1}{Department of Astrophysical Sciences, Princeton University,
Princeton, New Jersey 08544, USA}
\altaffiltext{2}{Max-Planck Institute for Extraterrestrial Physics,
Garching by Munich, 85748, Germany; 
munan@mpe.mpg.de}
\altaffiltext{2}{Center for Computational Astrophysics, Flatiron Institute, New York, NY, 10010}

\begin{abstract}
    $\CO(J=1-0)$ line emission is a widely used observational tracer of
    molecular gas, rendering essential the $X_\CO$ factor, which is applied to 
    convert $\CO$ luminosity to $\Ht$ mass. We use numerical simulations to 
    study how $X_\CO$ depends on numerical resolution,
    non-steady-state 
    chemistry, physical environment, and 
    observational beam size. Our study employs 3D magnetohydrodynamics (MHD)
    simulations of galactic disks with solar neighborhood conditions,
    where star formation
    and the three-phase interstellar medium (ISM) are self-consistently
    regulated by gravity and stellar feedback. Synthetic $\CO$ maps are
    obtained by post-processing the MHD simulations with
    chemistry and radiation transfer. We find that 
    $\CO$ is only an approximate tracer of $\Ht$. On parsec scales,
    $W_\CO$ is more fundamentally a measure of mass-weighted volume density, 
    rather than $\Ht$ column density.
    Nevertheless, 
    $\langle{X}_\CO\rangle=0.7-1.0\times10^{20}~\mr{cm^{-2}K^{-1}km^{-1}s}$ 
    consistent with
    observations, insensitive to the evolutionary ISM state
    or radiation field strength if
    steady-state
    chemistry is assumed. Due to non-steady-state
    chemistry, younger molecular clouds have 
    slightly lower $\langle{X}_\CO\rangle$ and flatter profiles of $X_\CO$ 
    versus extinction than older ones. The $\CO$-dark $\Ht$
    fraction is $26-79\%$, anti-correlated with the average
    extinction. As the observational beam size 
    increases from $1~\mathrm{pc}$ to $100~\mathrm{pc}$, $\langle{X}_\CO\rangle$ increases by 
    a factor of $\sim{2}$. 
    Under solar neighborhood conditions, $\langle{X}_\CO\rangle$ in molecular
        clouds is converged at a numerical resolution of $2~\mathrm{pc}$. However, the total $\CO$
    abundance and luminosity are not converged even at the numerical
    resolution of $1~\mathrm{pc}$.
    Our simulations successfully reproduce the
    observed variations of $X_\CO$ 
    on parsec scales, as well as the dependence of $X_\CO$ on extinction
    and the $\CO$ excitation temperature.
\end{abstract}

\section{Introduction}
Molecular clouds are the birth places of stars. In addition, molecular gas is
the dominant ISM component in dense and shielded environments.
Measuring the properties
of molecular clouds is therefore critical to understanding the ISM and 
star formation in the
Milky Way and beyond. However, the most abundant molecule in the ISM,
molecular hydrogen $\Ht$, is not directly observable in emission 
at typical ISM temperatures due to its low mass
and lack of dipole moment. As a result, the second most abundant molecule,
$\CO$, is often used as an observational tracer for $\Ht$. The standard
technique employs a conversion factor $X_\CO$ to relate the observed
velocity-integrated intensity of $\CO(J=1-0)$ line emission 
$W_\CO$ to the $\Ht$ column density $N_\Ht$, 
\begin{equation}
    N_\Ht = X_\CO W_\CO.
\end{equation}

Although the $\CO(J=1-0)$ line emission is bright and easy to detect 
with ground based radio telescopes, it is often very optically thick.
Many observational studies have measured $X_\CO$ by deriving the $\Ht$ mass
independently
of CO emission, via 
dust emission or extinction, gamma-ray emission, or the virial theorem
\citep[e.g.][]{Dame2001, Lombardi2006, SM1996, Solomon1987}. Surprisingly, the
value of $X_\CO$ only varies within a factor of $\sim 2$ for many molecular
clouds in the Milky Way and local disk galaxies. This has motivated the
adoption of a constant standard $X_\CO$ conversion factor in the literature, 
$X_\mr{CO, MW}=2\times 10^{20} ~\mr{cm^{-2}K^{-1}km^{-1}s}$ 
\citep[see review by][]{BWL2013}.

It is important to note that this standard $X_\CO$
is an average value for nearby molecular clouds on scales of tens of parsecs.
$X_\CO$ is empirically known to vary both on small scales, and for molecular clouds in
different environments.
One of the earliest studies of $X_\CO$, \citet{Solomon1987}, suggested that $X_\CO$
varies by a factor of a few for molecular clouds in the Milky Way, decreasing
with increasing $\CO$ luminosity. Recent high-resolution observations have found
that $X_\CO$ can vary by more than an order of magnitude on parsec scales, 
although the averages of $X_\CO$ over individual
molecular clouds are within a factor of $\sim 2$ of
the standard Milky Way value \citep{Pineda2008, Ripple2013, LeeMY2014,
Kong2015, Imara2015}.  
Beyond nearby molecular clouds, 
$X_\CO$ in the Galactic center is a factor of $\sim 4$ lower than
the mean value in the disk \citep{Blitz1985, Ackermann2012}, 
and similar results are found for the central regions in nearby spiral
galaxies \citep{Sandstrom2013}. 
High surface density starburst regions have $X_\CO$ significantly below
$X_\mr{CO, MW}$ \citep[e.g.][and references therein]{DS1998, BWL2013}.
Observations also indicate $X_\CO$ can be much 
higher than the standard Milky Way value in low metallicity
galaxies \citep{Israel1997, Leroy2011}. 

Theoretical models and numerical simulations have provided insights into the
$X_\CO$ conversion factor. 
\citet{Wolfire1993} constructed spherical cloud models with a photodissociation
region (PDR) code, and suggested that $X_\CO$ is only weakly dependent on
the incident far-ultraviolet (FUV) radiation field strength, and insensitive
to the small variations in metallicity up to a reduction of metallicity by a
factor of 5 relative to the solar neighborhood. These models rely on simple
assumptions about cloud structure and kinematics. 
To model molecular clouds with more realistic structure,
many numerical
simulations have been carried out to study 3D turbulent molecular clouds 
with self-consistent, time-dependent chemistry and radiation transfer
\citep[e.g.][]{GM2011, Shetty2011, Shetty2011b, GC2012a, Szucs2016}.
\citet{Shetty2011b} and \citet{Szucs2016} found
similar cloud-average 
$X_\CO$ to the standard observed value (with significant variations on
smaller-than-cloud scales).
\citet{Shetty2011b} concluded that
$X_\CO$ has a weak dependence on gas density, temperature, and velocity, 
and the nearly constant $X_\CO$ is the result of the limited range of
physical properties found in the nearby molecular clouds. 
However, these
simulations consider molecular clouds to be isolated from the large scale
galactic ISM, and their key physical properties such as the average
density and velocity dispersion are set artificially based on 
the initial conditions of the simulations and prescribed turbulent driving.

In recent years, more efforts have
been made to investigate $X_\CO$ in global galaxy simulations 
\citep{Narayanan2011, Narayanan2012, Feldmann2012, Duarte-Cabral2015}. 
With resolutions of tens of parsecs, however, global galaxy simulations
cannot resolve 
substructures in molecular clouds, and sub-grid models are generally required to
estimate the $\CO$ emission. There is no systematic study of the
dependence of $X_\CO$ on the numerical resolution in the literature.
Moreover, the comparisons between simulations
and observations are often focused on the cloud-average $X_\CO$.
Despite the rich observational data, little comparison has been made 
regarding to the variation of $X_\CO$ within molecular clouds on parsec
or smaller scales. Furthermore, as observations of galactic and extragalactic
molecular gas probe a range of scales, it is important to understand how
$X_\CO$ may vary with the effective area of a radio beam.

In this paper, we present a new study of the $X_\CO$ conversion factor in MHD
galactic disk simulations with solar neighborhood conditions and 
$1-4~\pc$ resolutions. The
high-density
clouds are
formed and destroyed self-consistently within the turbulent, multiphase, 
magnetized ISM by gravity and stellar feedback.
In our models, the distribution of
$\Ht$ and $\CO$ is obtained by post-processing the MHD simulations with chemistry
and radiation transfer.
While ideally all dynamics and chemistry would be self-consistent, 
\citet{GC2012a} pointed out that the gas temperature is not sensitive
to chemistry in the neutral ISM  \citep[see also][]{GOW2016}; as a consequence, dynamical simulations may
still represent ISM structure fairly accurately even if they do not include
time-dependent chemistry.

Using our models,
we investigate the dependence of $X_\CO$ on numerical resolution,
non-equilibrium (i.e.
non-steady-state) 
chemistry, variation in large-scale ISM structure and star
formation rates, and the observational beam size. 
Our analyses also identify the density and shielding conditions that are
required for $\Ht$ and $\CO$ formation (which differ significantly) in
realistic clouds, and break down the dependence of $W_\CO$ on microphysical
properties.
Additionally, we perform detailed comparisons with observations of $X_\CO$
in nearby molecular clouds at parsec scales.

The structure of this paper is as follows. In Section \ref{section:method}, we
describe the method of our simulations and the parameters in the
numerical models. In Section \ref{section:results}, we show our results and
comparisons with observations. The main findings of this work are summarized in
Section \ref{section:summary}.

\section{Method}\label{section:method}
To investigate the $X_\CO$ conversion factor in molecular clouds, we
carry out MHD simulations of galactic
disks, and post-process the results from MHD simulations with chemistry 
to obtain the distribution of molecular gas, including $\Ht$ and $\CO$.
Then we use
line radiation transfer code to model the $\CO$ emission from molecular clouds.

\subsection{MHD Simulation\label{section:MHD}}
The MHD simulation is performed with the {\sl TIGRESS}
(Three-phase Interstellar medium in Galaxies Resolving Evolution with 
Star formation and Supernova feedback) framework introduced by
\citet[][hereafter \citetalias{KO2017}]{KO2017}.
Here we briefly describe the key physics in the simulations, and refer
the readers to \citetalias{KO2017} for more extensive descriptions.

The {\sl TIGRESS} simulations model a kpc-sized region of a galactic disk where
the turbulent, multiphase, magnetized
ISM is self-consistently modeled with resolved star formation and feedback.
The physics are implemented within the {\sl Athena} code \citep{Stone2008}.
The ideal MHD equations are solved in a vertically-stratified local shearing box
\citep[e.g.][]{SG2010}. Self-gravity from gas and young stars are included by
solving Poisson's equation,
while a fixed vertical gravitational potential
represents  the old stellar disk and
the dark matter halo. Sink particles are implemented to represent star clusters,
and feedback from massive stars  
are included based on a population synthesis model 
\citep[STARBURST99;][]{Leitherer1999}. Both supernovae in star clusters
and from runaway OB stars are included.
The radiative heating and cooling of the gas are assumed to be optically
thin. The heating of cold and warm neutral gas
is from the photo-electric effect on dust grains; in the simulations the
heating rate is time dependent and scales with the instantaneous FUV luminosity
of the star cluster particles.
The cooling rate is obtained
from the local gas density and temperature using a simple cooling function
appropriate for the ionized and atomic ISM (combination of \citet{SD1993} and
\citet{KI2002}). 

The simulations self-consistently generate a representation of the
turbulent and magnetized 
three-phase ISM. In the fiducial model with solar neighborhood parameters, 
much of the volume is occupied by hot ionized gas, and
most of the mass near the midplane 
is in the warm and cold neutral medium (WNM and CNM), 
similar to the observed ISM in
the Milky Way and nearby galaxies. Although molecular gas is not explicitly
modeled in the {\sl TIGRESS} simulations, large structures of dense gas
naturally develop, and in reality molecular gas
would form within the regions of the CNM where the gas is 
dense and shielded. We model the
formation of molecular gas by post-processing the {\sl TIGRESS} simulations with
chemistry and shielding, which is described in detail 
in Section \ref{section:post-processing}.

We adopt the fiducial solar neighborhood model of \citetalias{KO2017}. 
The simulation domain size is
$L_x=L_y=1024~\pc$ and $L_z=4096~\pc$. The initial gas surface density
$\Sigma = 13~M_\odot \pc^{-2}$. The simulation reaches a quasi-steady
state after $t\approx 200~\Myr$. The total mass of the gas in the
simulation slowly declines as the gas turns into stars or leaves the
simulation domain as galactic winds. In this paper, we focus on the simulation
during the time frame $t=350-420~\Myr$ when the surface density of the gas is
in the range $9~M_\odot \pc^{-2} < \Sigma < 10~M_\odot \pc^{-2}$.

In order to study the effect of numerical resolution on $X_\CO$, we
consider the simulation with three different resolutions: $\Delta x=4$, $2$, and
$1~\pc$. The $4~\pc$ simulation starts from $t=0$ with the initial condition
described in \citetalias{KO2017}, and runs until $t=700~\Myr$. 
To save computational time, we use
an ``extraction'' method to refine the resolution. We use the
output of the $4~\pc$ simulation at time $t=350~\Myr$ as the initial condition
of the $2~\pc$ simulation, and run that for $70~\Myr$ (until $t=420~\Myr$).
Similarly, we use the output of the $2~\pc$ simulation at $t=378~\Myr$ as the
initial condition of the $1~\pc$ simulation, and run that for $4~\Myr$ (until
$t=382~\Myr$). We also reduce the domain size in the z-direction to $L_z=2240~\pc$
for the $2~\pc$ simulation and to $L_z=896~\pc$ for the $1~\pc$ simulation. Because
the scale-hight $H\sim 100~\pc$ for the CNM and $H\sim 400~\pc$ for the WNM, the
simulation domain in the z-direction is big enough
to capture most of the mass in the neutral and
molecular ISM.

When refining from a coarser resolution, it takes some time for
the turbulence to cascade down to smaller scales and create finer structures.
The line-width size relation \citep[e.g.][]{Larson1981, Solomon1987, HB2004,
HD2015},
\begin{equation}\label{eq:v_l}
    v(l) \sim 0.7~\mr{km/s}~\left( \frac{l}{\mr{pc}} \right)^{1/2},
\end{equation}
gives the expected timescale for turbulent cascade
in the dense ISM:
\begin{equation}\label{eq:t_turb}
    t_\mr{turb}(l)=\frac{l}{v(l)} = 1.4~\Myr~( l/\mr{pc} )^{1/2}.
\end{equation}
We only use
the outputs from the $2~\pc$ and $1~\pc$ simulations $4-6~\Myr$ after the
extraction from coarser resolution, allowing sufficient time for the turbulence
to develop at the refined resolution.

The density threshold for sink
particle creation, $n_\mr{thr}$, also depends on the resolution of the
simulation. A sink particle is created if the cell is at the local
gravitational potential minimum, the flow is converging, and the density of the
cell exceeds the Larson-Penston threshold \citep{Larson1969, Penston1969}
suggested by \citet{GO2013},
\begin{equation}
    \rho_\mr{thr} \equiv \rho_\mr{LP}(\Delta x/2) = \frac{8.86}{\pi}
    \frac{c_s^2}{G \Delta x^2}.
\end{equation}
The typical density threshold at the equilibrium CNM temperature is 
$n_\mr{thr}=2956$, $927$, and $304~\mr{cm^{-3}}$ for resolutions
$\Delta x=1$, $2$ and $4~\pc$.\footnote{This is
    assuming the heating rate of the CNM to be the solar neighborhood value
    $\Gamma = \Gamma_0 = 2\times 10^{-26}~\mr{erg~s^{-1}}$ 
    (\citetalias{KO2017} Equation (8)). However, 
    $n_\mr{thr}$ is insensitive to the change of $\Gamma$: 
    $n_\mr{thr}$ increases by less then a factor of
two when $\Gamma$ increases by a factor of ten.}

\subsection{Post-processing chemistry\label{section:post-processing}}
To model the chemical composition of the gas, we have developed a
post-processing module within the code {\sl Athena++} \citep{White2016}.
This module reads the output from {\sl TIGRESS} simulations and performs
chemistry calculations assuming the density and velocity in each
grid cell is fixed. 
We use the simplified chemical network of \citet{GOW2016}, which focuses
on the hydrogen, carbon, and oxygen chemistry, and gives accurate 
abundances of $\Ht$ and $\CO$. 
We assume an initial chemical composition of neutral atomic gas,
with all hydrogen in the form of $\mr{H}$, all carbon in $\mr{C}$, 
all oxygen in $\mr{O}$, and all silicon in 
$\mr{Si}$. The initial temperature is the same as the output from
MHD simulations. Then we evolve the chemistry, temperature, and radiation field
(see below) for time $t_\mr{chem}=50~\Myr$, 
so that the chemical abundances of the gas reach
steady state. In other words, we do not self-consistently calculate the
time-dependent gas dynamics and chemistry, but instead 
consider the state in which the
chemistry and temperature have reached a equilibrium, consistent
  with radiative heating and ISM structure
as determined by the MHD simulations.
Because gas cooling is not sensitive to the chemical composition, chemistry
has minimal effect on the gas dynamics \citep{GC2012a, GOW2016}. However, dust
shielding can reduce the gas heating, and lower the gas temperature by a factor
of $\sim 2$ in shielded regions of the CNM where molecular gas
forms.
\footnote{This typical reduction in temperature
in high density regions ($n_\mr{H} \gtrsim 10~\mr{cm^{-3}}$)
is found by comparing the initial
temperature output from the MHD simulation 
to the steady-state temperature from
the post-processing chemistry simulation.
}
In return, gas dynamics can also influence the chemical composition.
For example, the timescale for $\Ht$ formation can be longer than the 
turbulent crossing time in the molecular clouds, which may lead to much lower
$\Ht$ abundance than the equilibrium
values \citep{GOW2016}.  
The temporal dependence of the chemical state and observable CO
  properties are considered in 
Section \ref{section:non-equilibrium}.

The heating and cooling of the gas is calculated simultaneously with chemistry,
with the details described in \citet{GOW2016}. We slightly modify the
parameter $\tilde{N} (\CO)$ for $\CO$ cooling in \citet{GOW2016} by setting
\begin{equation}\label{eq:til_NCO}
    \tilde{N} (\CO) = \frac{n(\CO)}{\mr{max} \left(\langle|\di v/\di r|\rangle,
    v_\mr{th}/l_\mr{esc}\right)},
\end{equation}
where $\langle |\di v/\di r| \rangle$  is the mean (absolute) velocity gradient
across the six faces of each grid cell in the simulation,
$v_\mr{th}=\sqrt{2kT/m(\CO)}$ the thermal velocity of $\CO$ molecules, and
$l_\mr{esc}=100~\pc$ the maximum length scale for a photon to escape. Using the
maximum of two terms in the denominator of Equation (\ref{eq:til_NCO}) ensures
that there is a minimum probability for the photon to escape when the local
velocity gradient is small, given a maximal molecular cloud size,
$\lesssim 100~\pc$. This
formalism is consistent with the large velocity gradient (LVG) and escape
probability approximation we adopted in carrying out the synthetic observations 
of $\CO$ line emission (Section \ref{section:synthetic_obs}). 

In order to compute the photoionization and photodissociation rates in the chemistry
network, a radiation transfer scheme is needed to calculate the reduction of
FUV radiation by dust and molecule shielding. We use the six-ray approximation
\citep{NL1997, NL1999, GM2007}: in each cell, the radiation field is 
calculated by ray-tracing and averaged over six directions along the Cartesian
axes. The incident radiation field is assumed to come from the edge of the
computational domain
along each ray, and has the initial intensity the same as that in the
MHD simulations (the MHD simulations themselves do not include shielding).
The main advantage of this approach is the low computational cost. When
comparing to ray-tracing along many more different angles, the six-ray
approximation gives reasonably accurate results \citep{Safranek-Shrader2017}.
Because chemistry and radiation depend on each other, we iterate to solve
the chemistry equations and six-ray radiation transfer.

\subsection{Synthetic Observation of $\CO$ Line Emission 
\label{section:synthetic_obs}}
To model the $\CO(J=1-0)$ line emission, we apply the publicly available 
radiation transfer code {\sl RADMC-3D} \citep{Dullemond2012} with
chemistry and temperature obtained as described in Section
\ref{section:post-processing}.\footnote{We set a temperature ceiling
of $T=200~\mr{K}$ for the temperature input, because $\CO$ only forms within
the CNM where $T\lesssim 100~\mr{K}$, and a high temperature input 
from the WNM and hot gas introduces additional
computational cost for calculating the $\CO$ population levels in regions 
where the $\CO$ abundance is essentially zero. We have tested
using a higher temperature ceiling of $1000~\mr{K}$ and confirmed that it gives
the same result.}
We select the mid-plane region $|z| < 256~\pc$,
where almost all molecules are found. 
$\Ht$ is assumed to be the only collisional partner with
$\CO$, and we use a fixed ortho-to-para ratio of 3:1.\footnote{The collisional
coefficients for ortho- and para- $\Ht$ are very similar, and we have tested
that a ortho-to-para ratio of 1:1 gives very similar results.} The synthetic
observations are performed along the z-axis, i.e., the
observer is looking at the galactic disk face-on. This avoids cloud blending, as
all molecular clouds form near the mid-plane of the galactic disk. 

The $\CO$ population levels are calculated by using the LVG and
escape probability approximation, which is implemented in {\sl RADMC-3D} by
\citet{Shetty2011}. This approximation allows the population levels to be
calculated locally in each cell. The escape probability is
\begin{equation}\label{eq:beta}
\beta = \frac{1-\mr{e}^{-\tau}}{\tau},
\end{equation}
and the optical depth $\tau = \mr{min}\left( \tau_\mr{LVG}, \tau_\mr{EscProb}
\right)$. The LVG approximation gives
\begin{equation}\label{eq:tau_LVG}
    \tau_\mr{LVG} = \frac{\lambda_{10}^3}{8\pi} \frac{A_{10}n_\CO}{
    \langle |\di v/\di r| \rangle}
    f_1 \left( \frac{f_0/g_0}{f_1/g_1} - 1 \right),
\end{equation}
where $A_{10}$ is the Einstein A coefficient 
$A_{10}=7.203\times 10^{-8}~\mr{s^{-1}}$, $n_\CO$ the number density of $\CO$
molecules, $g_0=1$ and $g_1=3$ the degeneracy for $J=0$ and $J=1$ levels, 
$f_0=n_0/n_\CO$ and $f_1 = n_1/n_\CO$ the fraction of $\CO$ molecules in
$J=0$ and $J=1$ levels, where $n_0$ and $n_1$ are the level populations, 
$\langle |\di v/\di r| \rangle$ the same as that in Equation
(\ref{eq:til_NCO}).
The optical depth from the escape probability approximation
is set by a typical length-scale $L_\mr{EscProb}$, and can be written in the
same from as Equation (\ref{eq:tau_LVG}) by substituting $\langle |\di v/\di r|
\rangle$ with $\sqrt{\pi}v_\mr{tot}/L_\mr{EscProb}$ \citep{Draine2011}.
Here $v_\mr{tot}$ is the total velocity dispersion (see below).
We adopt $L_\mr{EscProb}=100~\pc$, consistent with the
 $\CO$ line cooling in Equation (\ref{eq:til_NCO}). In our
simulation, the velocity gradient is usually relatively large, and in most
cells $\tau=\tau_\mr{LVG}$.

Ray-tracing is performed after the $\CO$ level populations are obtained. 
In general, the emission line intensity is determined by radiative
transfer \citep[e.g.][]{Draine2011}:
\begin{equation}\label{eq:dInu}
    \di I_\nu = - I_\nu \di \tau_\nu + S_\nu \di \tau_\nu,
\end{equation}
where $I_\nu$ is the line intensity at frequency $\nu$, $S_\nu$ the source
function, and $\tau_\nu$ the optical depth. $\tau_\nu$ depends on the line
profile, which is set by the velocity dispersion
$v_\mr{tot} = \sqrt{v_\mr{th}^2 +
v_\mr{turb}^2}$. We include a sub-grid 
``micro-turbulent'' velocity dispersion according
to the line-width size relation (Equation (\ref{eq:v_l})),
\begin{equation}\label{eq:v_turb}
v_\mr{turb} = 0.7~\mr{km/s}~\left( \frac{\Delta x}{\mr{pc}} \right)^{1/2},
\end{equation}
where $\Delta x$ is the resolution of the simulation. We also include a
background blackbody radiation field with temperature $T_\mr{CMB}=2.73~\mr{K}$ 
from the cosmic microwave background (CMB).

We run {\sl RADMC-3D} with a
passband from $-20~\mr{km/s}$ to $20~\mr{km/s}$ (wide enough to include all
$\CO$ emission) and velocity resolution of
$0.5~\mr{km/s}$. RADMC-3D produces spectral position-position-velocity (PPV) cubes of the
$\CO(J=1-0)$ line.
We then interpolate $I_\nu$ to a finer velocity resolution of
$0.07~\mr{km/s}$, and calculate the total $\CO(J=1-0)$ line intensity in each
observed pixel, $W_\CO$, by integrating $I_\nu$ over
all velocity channels that have emission above the detection limit,
$T_\mr{det}=0.4~\mr{K}$.  This approach matches the typical
velocity resolution and sensitivity in observations of nearby molecular
clouds \citep[e.g.][, see also Table \ref{table:XCO_obs}]{
Ridge2006, Pineda2008, Pineda2010, Ripple2013, Lee2014}.
 We define the
 ``$\CO$-bright'' region as pixels with
 $W_\CO > 0.1 ~\mr{K}~\mr{km/s}$, and calculate
$X_\CO$ for each pixel in the $\CO$-bright region. The average $X_\CO$,
$\langle X_\CO \rangle = \sum N_\Ht/\sum W_\CO$ is also calculated only within
the $\CO$-bright region, similar to the common approach in observations
\citep[e.g.][]{Pineda2008, Ripple2013}. We define the fraction of $\CO$-dark $\Ht$,
\begin{equation}\label{eq:f_dark}
    f_\mr{dark}\equiv \frac{M_\Ht(W_\CO<0.1~\mr{K~km~s^{-1}})}{M_\mr{H_2,tot}}.
\end{equation}

\subsection{The beam size in synthetic
observations\label{section:method:beamsize}}
The default beam size in our synthetic observations is the same as the
numerical resolution in the MHD simulations. In real observations, the beam size (in
physical units) varies depending on the telescope and the distance of the
object. The dust extinction or emission map used to derive $\Ht$ column
densities typically has coarser resolution than the
$\CO$ map. To analyze the $X_\CO$ values, the dust map and $\CO$ map
are smoothed to a common resolution (usually the resolution of the dust map),
which we refer to as the ``beam size''.\footnote{Note that this is often
called ``pixel size'' in observations. We use ``beam size'' to distinguish from the
``pixel size'' determined by the numerical resolution of our simulation and
synthetic radiative transfer grid.}
The velocity resolution and sensitivity also vary in observations. We have
compiled the observational parameters from the literature of $X_\CO$ observations
in the Milky Way and nearby galaxies in Table \ref{table:XCO_obs}. All the
observations listed used $\Ht$ mass estimation from dust extinction or
emission. We also list $\langle X_\CO \rangle$ obtained by the observations
when available. 

\begin{table*}[htbp]
    \caption{Observational parameters in selected $X_\CO$ literature}
    \label{table:XCO_obs}
    \begin{tabular}{cccc cccc}
        \tableline
        \tableline
        reference &beam size &$\CO$ map res.
        &distance\tablenotemark{a}
        &object &velocity res. ($\mr{km/s}$)
        &$T_\mr{det}(\mr{K})$\tablenotemark{b} 
        &$\langle X_\CO \rangle_{20}$\\
        \tableline
        \citet{Ripple2013} &0.2 pc &0.1 pc &420 pc
        &Orion   &0.2   &2    &$1.4$ \\
        \citet{LeeMY2014}  &0.36 pc &0.06 pc &280 pc
        &Perseus &0.064 &0.8 &$0.3$\tablenotemark{c} \\
        \citet{Pineda2008} &0.4 pc &0.06 pc &280 pc
        &Perseus &0.064 &0.35 &$2\pm 1$ \\
        \citet{Leroy2011}  &60 pc  &5.8 pc &50 kpc
        &LMC     &0.1   &0.35 &$3.0$\\
        \citet{Leroy2016}  &60 pc  &11-60 pc  &0.05-21.5 Mpc
        &nearby galaxies\tablenotemark{d} &1.6-5  &0.03-0.2   &--\\
        \citet{Smith2012}  &140 pc &90 pc &780 kpc
        &M31     &2.6   &0.03 &$1.9\pm 0.4$ \\
        \citet{Sandstrom2013}\tablenotemark{e} &0.6-4 kpc &0.2-1.2 kpc &3.6-21.4 Mpc
        &spiral galaxies &2.6 &0.02-0.04 &$1.4-1.8$\tablenotemark{f} \\
        \tableline
        \tableline
    \end{tabular}
    \tablenotetext{1}{Distance of Perseus and Orion molecular clouds are taken
    from \citet{Schlafly2014}.}
    \tablenotetext{2}{Detection limit for $\CO(J=1-0)$ line emission. Same as
    the mean RMS noise per velocity channel in observations.}
    \tablenotetext{3}{Note that the $\langle X_\CO \rangle$ in \citet{LeeMY2014} is smaller
    than that determined by \citet{Pineda2008}. \citet{LeeMY2014} states that
    the discrepancy mainly results from different adopted dust-to-gas ratio
    and the consideration of $\mr{HI}$ gas.}
    \tablenotetext{4}{Antennae, LMC, M31, M33, M51, and M74.}
    \tablenotetext{5}{Observations used the $\CO(J=2-1)$ line and assumed
        a fixed line ratio (2-1)/(1-0)=0.7.}
    \tablenotetext{6}{This is the average $\langle X_\CO \rangle $ in 
        low-inclination galaxies. The dispersion is about 0.3 dex.} 
\end{table*}

We investigate the effect of beam size on $X_\CO$ in Section
\ref{section:result:beamsize}. The adopted parameters and beam sizes
are listed in Table \ref{table:para_obs}, which is
designed to match the typical observational parameters listed in Table
\ref{table:XCO_obs}.
The synthetic observations with default beam size are based on the original
model data (the same as the numerical resolution). To create synthetic maps
with larger effective beam, we first smooth out the PPV cubes
produced by RADMC-3D to the desired $\CO$ map resolution. Then we match the
corresponding velocity resolution from the default 
$0.5~\mr{km/s}$ in the PPV cubes, by either interpolating to finer or
integrating
to coarser velocity resolution. We integrate over all velocity channels with
emission above the detection limit $T_\mr{det}$ and obtain a 2D map of $W_\CO$ at
the corresponding $\CO$ map resolution. Then both the map for $A_V(N_\Ht)$ and
the map for $W_\CO$ are smoothed to the common resolution of the beam size, for
which $X_\CO$ is calculated. We note
that the ``beam'' is square, not circular.
\footnote{We have compared
    results for our square beam
    to the result for a circular gaussian beam, and find that it makes very little difference for
$\langle X_\CO \rangle$.} 
The $\CO$-bright region, for
which $\langle X_\CO \rangle$ is calculated, is defined as pixels with 
$W_\CO > 3 T_\mr{det} \Delta v$, where
$\Delta v$ is the width of the velocity channel. 

\begin{table}[htbp]
    \caption{Parameters for synthetic observations}
    \label{table:para_obs}
    \begin{tabular}{cccc}
        \tableline
        \tableline
        beam size(pc) &$\CO$ map res.(pc) &velocity res.($\mr{kms}$) &$T_\mr{det}(\mr{K})$\\
        \tableline
        1      &1    &0.07    &0.4 \\
        2      &2    &0.07    &0.4 \\
        4      &2    &0.07    &0.4 \\
        8      &2    &0.07    &0.4 \\
        16     &2    &0.07    &0.4 \\
        32     &2    &0.07    &0.4 \\
        64     &4    &0.1     &0.35 \\
        128    &64   &2.6     &0.03 \\
        512    &128  &2.6     &0.03 \\
        1024   &256  &2.6     &0.03 \\
        \tableline
        \tableline
    \end{tabular}
\end{table}

\subsection{Model parameters}
We consider three sets of models designed to study different conditions
that may affect $X_\CO$: the numerical resolution, non-equilibrium
chemistry,
and variation in the galactic environment (ISM structure and ambient radiation
field). The parameters for our models are summarized in Table \ref{table:model}. 
Model names denote changes in numerical resolution (RES-1pc, etc.), chemical
evolution time (TCHEM-5Myr, etc.), and simulation snapshot time (T-356Myr,
etc.). Note
that $t$ is the time for the MHD simulation, and $t_\mr{chem}$ is the time for
the post-processing chemistry, as detailed in Section \ref{section:post-processing}.
RES-1pc and TCHEM-50Myr are two names for the same model, used for clarity 
in different sections discussing the numerical resolution or
evolving 
chemistry. To do a controlled study, we set the incident
radiation field strength $\chi=1$ (in \citet{Draine1978} units, corresponding
to $J_\mr{FUV} = 2.7 \times 10^{-3} \mr{erg~cm^{-2} s^{-1}}$)
for all models that intercompare numerical resolution and
non-equilibrium
chemistry (model IDs starting with RES or TCHEM).
In the set of models for studying the variation in galactic environments 
(model IDs starting with T), $\chi$
is obtained from the star cluster particles as described in \citetalias{KO2017}.

\begin{table}[htbp]
    \caption{Model parameters}
    \label{table:model}
    \begin{tabular}{cccc}
        \tableline
        \tableline
        ID &Resolution ($\pc$) & $t~(\Myr)$ & $t_\mr{chem}~(\Myr)$\\ 
        \tableline
        \multicolumn{4}{l}{Convergence of numerical resolution:}\\
        RES-1pc &1 &382 &50\\
        RES-2pc &2 &382 &50\\
        RES-4pc &4 &382 &50\\
        \tableline
        \multicolumn{4}{l}{Non-equilibrium chemistry:}\\
        TCHEM-5Myr &1 &382 &5\\
        $\vdots$ & & &\\
        TCHEM-50Myr &1 &382 &50\\
        \tableline
        \multicolumn{4}{l}{Variation in galactic environments:}\\
        T-356Myr &2 &356 &50\\
        $\vdots$ & & &\\
        T-416Myr &2 &416 &50\\
        \tableline
        \tableline
    \end{tabular}
\end{table}

\section{Results}\label{section:results}
\subsection{Convergence study: effects of numerical resolution\label{section:result:res}}
\begin{figure*}[htbp]
\centering
\includegraphics[width=0.95\linewidth]{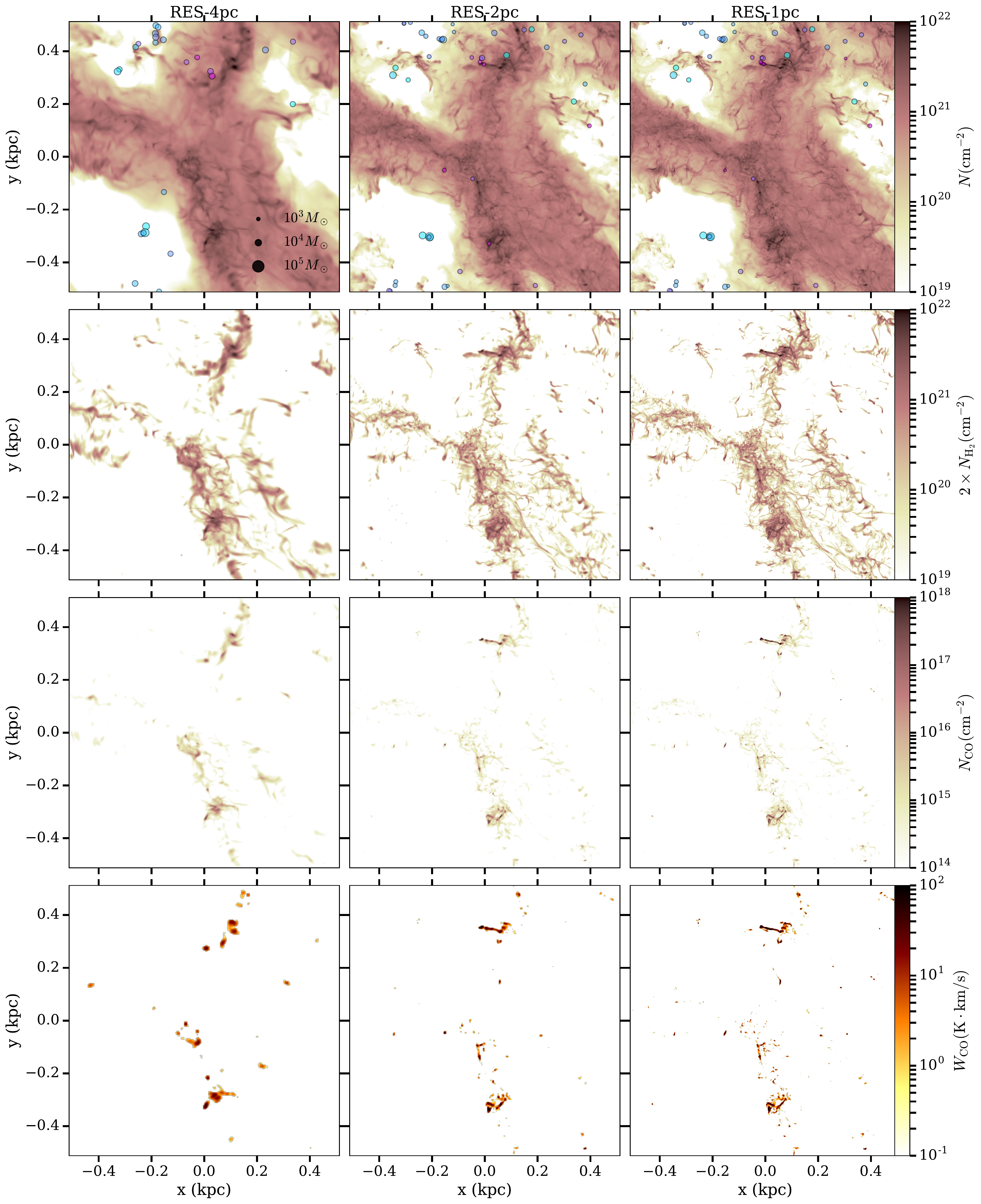}
\caption{The column density of all gas ($N$, {\sl first row}), molecular gas
    ($N_\Ht$, {\sl second row}), $\CO$ ($N_\CO$, {\sl third row}), and the intensity of the 
    $\CO(J=1-0)$ line ($W_\CO$, {\sl last row}) in models RES-4pc, ({\sl
    left}), RES-2pc ({\sl middle}) and RES-1pc ({\sl right}). 
    The young (age $<40~\Myr$) star
    clusters/sink particles formed in the simulations are shown as filled
    circles in the first row with $N$.  The area of the circles are proportional to
    the square root of the cluster masses, ranging from $10^3~M_\odot$ to
    $10^5~M_\odot$ (see legends in the top left panel), and the color
of the circles indicates the cluster age, from $0$ (magenta) to $40~\Myr$ (blue).}
\label{fig:res_overview}
\end{figure*}

In this section, we investigate the effect of numerical resolution on both
chemistry and $X_\CO$. An overview of the models RES-4pc, RES-2pc and RES-1pc
is shown in Figure \ref{fig:res_overview}, and the overall properties of the
models are listed in Table \ref{table:res_comparison}. As the
resolution increases, more small structures and dense gas forms in the
simulations. The locations of molecular clouds are similar in all three models,
but the small scale filamentary structures within the molecular clouds can only be
resolved in RES-2pc and RES-1pc. As we shall show (Section \ref{section:res_XCO}), 
at least $2~\pc$ resolution is needed 
to accurately determine the average $X_\CO$ in molecular clouds
for the Solar neighborhood conditions of the present simulations.

\begin{table*}[htbp]
    \caption{Overall properties of models for comparisons in numerical resolution and
    non-equilibrium chemistry}
    \label{table:res_comparison}
    \begin{tabular}{c ccc ccccc}
        \tableline
        \tableline
        model ID  &$M_\mr{tot}(M_\odot)$\tablenotemark{a}
        &$M_\Ht(M_\odot)$ &$M_\CO(M_\odot)$
        &$L_\CO~(\mr{K~km~s^{-1}pc^2})$\tablenotemark{b}
        &$\langle X_\CO \rangle_{20}$\tablenotemark{c}
        &$f_\mr{dark}$\tablenotemark{d}
        &$f_{100}$\tablenotemark{e} &$2 \langle f_{\Ht} \rangle$\tablenotemark{f}
        \\ 
        \tableline
        RES-4pc &$7.48\times 10^6$ &$5.76\times 10^5$ &$4.82\times 10^1$
        &$7.63\times 10^4$ &1.45  &69\% &0.4\% &11\%\\
        RES-2pc &$7.41\times 10^6$ &$5.55\times 10^5$ &$5.49\times 10^1$
        &$8.27\times 10^4$ &1.07  &75\% &0.9\% &10\%\\
        RES-1pc (TCHEM-50Myr) &$7.41\times 10^6$ &$6.89\times 10^5$ &$1.21\times 10^2$
        &$1.22\times 10^5$ &1.02  &71\% &2.3\% &13\%\\
        TCHEM-5Myr &$7.41\times 10^6$ &$2.46\times 10^5$ &$8.96\times 10^1$
        &$9.06\times 10^4$ &0.56  &67\% &2.3\% &5\%\\
        \tableline
        \tableline
    \end{tabular}
    \tablenotetext{1}{Total mass $M_\mr{tot} = 1.4 m_\mr{H} \int n \di V$. The
    factor 1.4 is from the helium abundance $f_\mr{He}=0.1$.}
    \tablenotetext{2}{Total luminosity of $\CO(J=1-0)$ line.}
    \tablenotetext{3}{Average $X_\CO$ in $\CO$-bright regions. 
    $\langle X_\CO \rangle_{20}=\langle X_\CO \rangle
    /(10^{20} \mr{cm^{-2}K^{-1}km^{-1}s})$.}
    \tablenotetext{4}{$\CO$-dark $\Ht$ gas fraction (see Equation
    (\ref{eq:f_dark})).}
    \tablenotetext{5}{Fraction of mass with density
    $n>100~\mr{cm^{-3}}$.}
    \tablenotetext{6}{Fraction of hydrogen in $\Ht$: 
    $2 \langle f_{\Ht} \rangle=M_\Ht/(M_\Ht + M_\mr{H})$. }
\end{table*}

\subsubsection{Molecular Abundances and Dependence of Chemistry on numerical
resolution\label{section:res_chemistry}}
\begin{figure*}[htbp]
\centering
\includegraphics[width=0.9\linewidth]{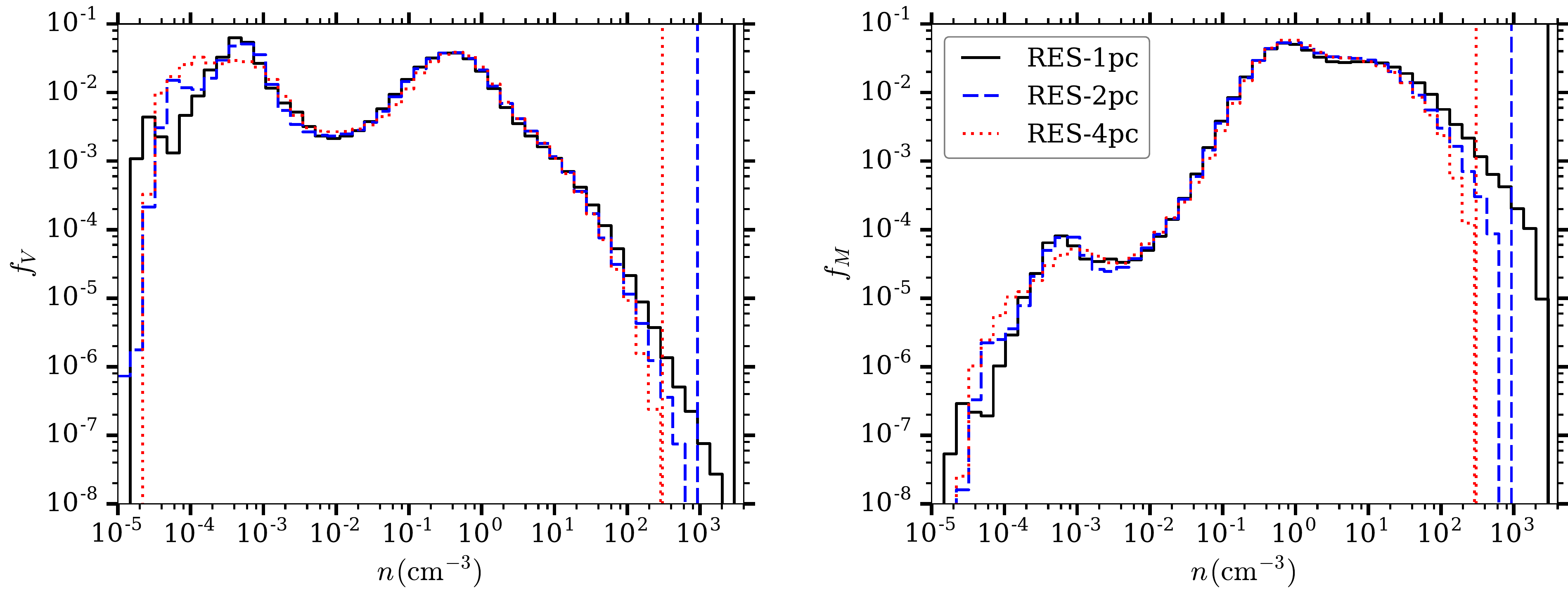}
\caption{Histograms of volume-weighted ({\sl left}) and mass-weighted 
    ({\sl right}) density $n$ in models RES-1pc (solid black), RES-2pc
    (dashed blue) and RES-4pc (dotted red). The $y$-axes are normalized 
    to show the fraction of volume $f_V$ or mass $f_M$ in each density bin.
    The vertical lines indicate the density threshold for sink
particle creation at the corresponding resolution in each model (Section
\ref{section:MHD}).}
\label{fig:nH_hist}
\end{figure*}

As the numerical resolution increases from $4~\pc$ to $1~\pc$, 
a larger fraction of mass in the simulations is in
the dense gas. This is quantified by the increase of $f_{100}$ (the fraction of
gas with density $n>100~\mr{cm^{-3}}$) with resolution in
Table \ref{table:res_comparison}, 
and the density distributions in Figure \ref{fig:nH_hist}.
The density distributions are similar at low densities where the gas is well
resolved.  At high densities, the distribution cuts off near the
density threshold for sink particle creation, where the unresolved dense gas
is converted into sink particles in the simulations.
As resolution increases, the density threshold for sink particle creation 
also increases, allowing denser gas to form.

\begin{figure*}[htbp]
\centering
\includegraphics[width=0.9\linewidth]{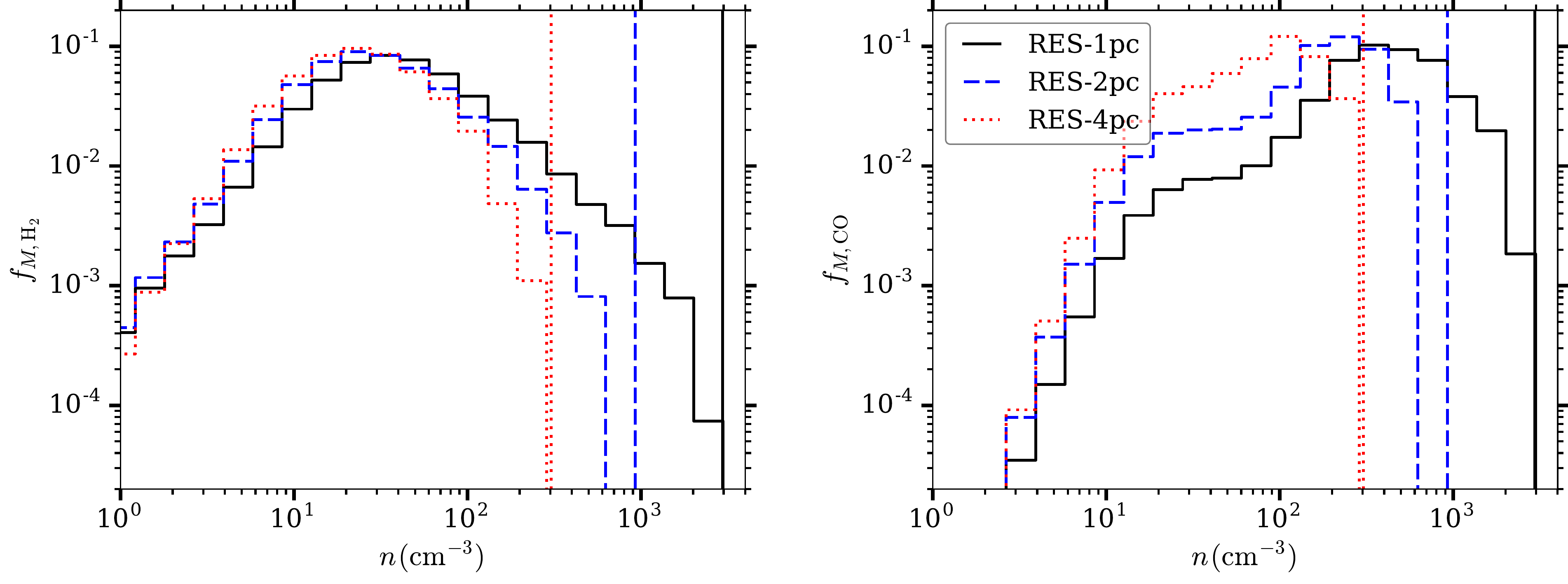}
\caption{Histograms of density, similar to Figure \ref{fig:nH_hist}, but
    weighted by $\Ht$ mass ({\sl left}) and $\CO$ mass ({\sl right}) in each cell.}
\label{fig:nH_hist_MH2_MCO}
\end{figure*}

The change of density distribution with resolution affects the chemical
compositions of the gas. As the resolution increases from $4~\pc$ to $1~\pc$,
the total $\Ht$ mass stays nearly constant, but the total $\CO$ mass increases
by a factor of nearly 3 (Table \ref{table:res_comparison}).\footnote{
In Table \ref{table:res_comparison}, $M_\Ht$ first decreases slightly
when the resolution increases from 4 pc to 2 pc, then increases again at 1 pc
resolution. This non-linear variation of $M_\Ht$ with resolution is actually a
result of temporal variations in the simulations. Because the supernova feedback
from the sink/cluster particles is stochastic,
simulations with the same initial condition
can develop slightly different density structures over time. We compared $M_\Ht$ and
$M_\CO$ in models RES-4pc and RES-2pc between the time when they have the same
initial condition (350 Myr) and the time of comparision in Table
\ref{table:res_comparison} (382 Myr). We found that the $\Ht$ mass in both
models are similar (up to $\sim 20\%$ variations), but the $\CO$ mass increases
significantly (up to a factor of $\sim 3$) in the RES-2pc model. The $\Ht$ and
$\CO$ mass weighted density histograms at different times also show very
similar features to Figure \ref{fig:nH_hist_MH2_MCO}. Therefore, the
conclusion from Figure \ref{fig:nH_hist_MH2_MCO} is robust despite the temporal
variations. }
The reason for this is evident in Figure \ref{fig:nH_hist_MH2_MCO}:
most $\Ht$ forms in
the density range of $n=10-100~\mr{cm^{-3}}$, which is already well resolved with
$4~\pc$ resolution. However,
most $\CO$ forms at $n \gtrsim 200~\mr{cm^{-3}}$, which is not well resolved with
$2~\pc$, maybe even $1~\pc$ resolution. Using adaptive mesh refinement (AMR)
models, \citet{Seifried2017} found a resolution of $\sim 0.2~\pc$ is needed
for the $\CO$ abundance to converge.

\begin{figure*}[htbp]
\centering
\includegraphics[width=0.9\linewidth]{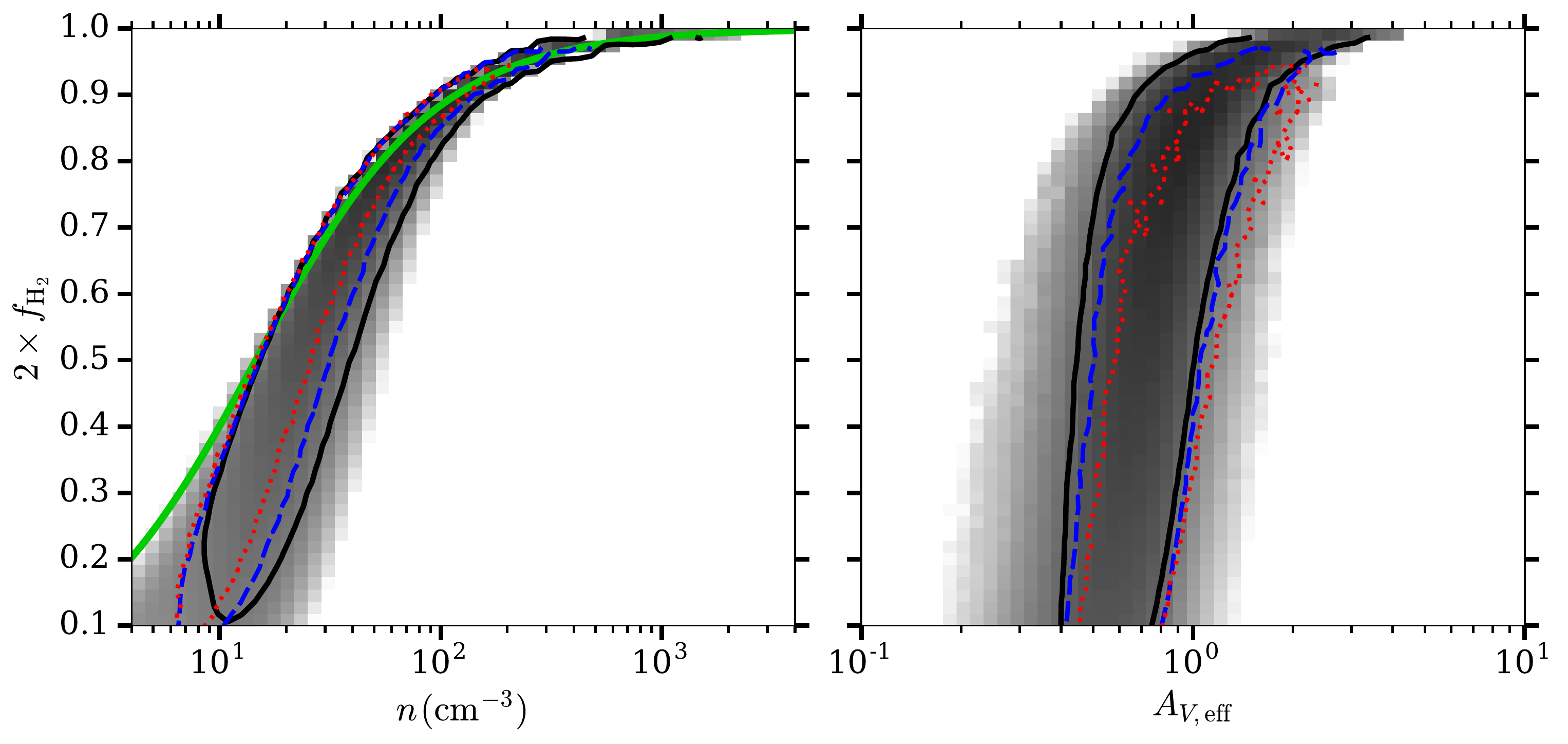}
\caption{Distributions of the $\Ht$ abundance $f_\Ht$ versus the gas density $n$
    ({\sl left}) and the effective extinction $A_{V, \mr{eff}}$ ({\sl right}). 
    The color scale shows the log of the $\Ht$ mass
    in each bin for model RES-1pc, spanning three orders of magnitude.
    The contours indicate 90\% of the $\Ht$ mass in models RES-1pc (black
    solid), RES-2pc (blue dashed), and RES-4pc (red dotted). The green line shows the
equilibrium $\Ht$ abundance assuming the FUV radiation is completely shielded 
(Equation (\ref{eq:H2_eq})).}
\label{fig:fH2_n_AVsh}
\end{figure*}

\begin{figure*}[htbp]
\centering
\includegraphics[width=0.9\linewidth]{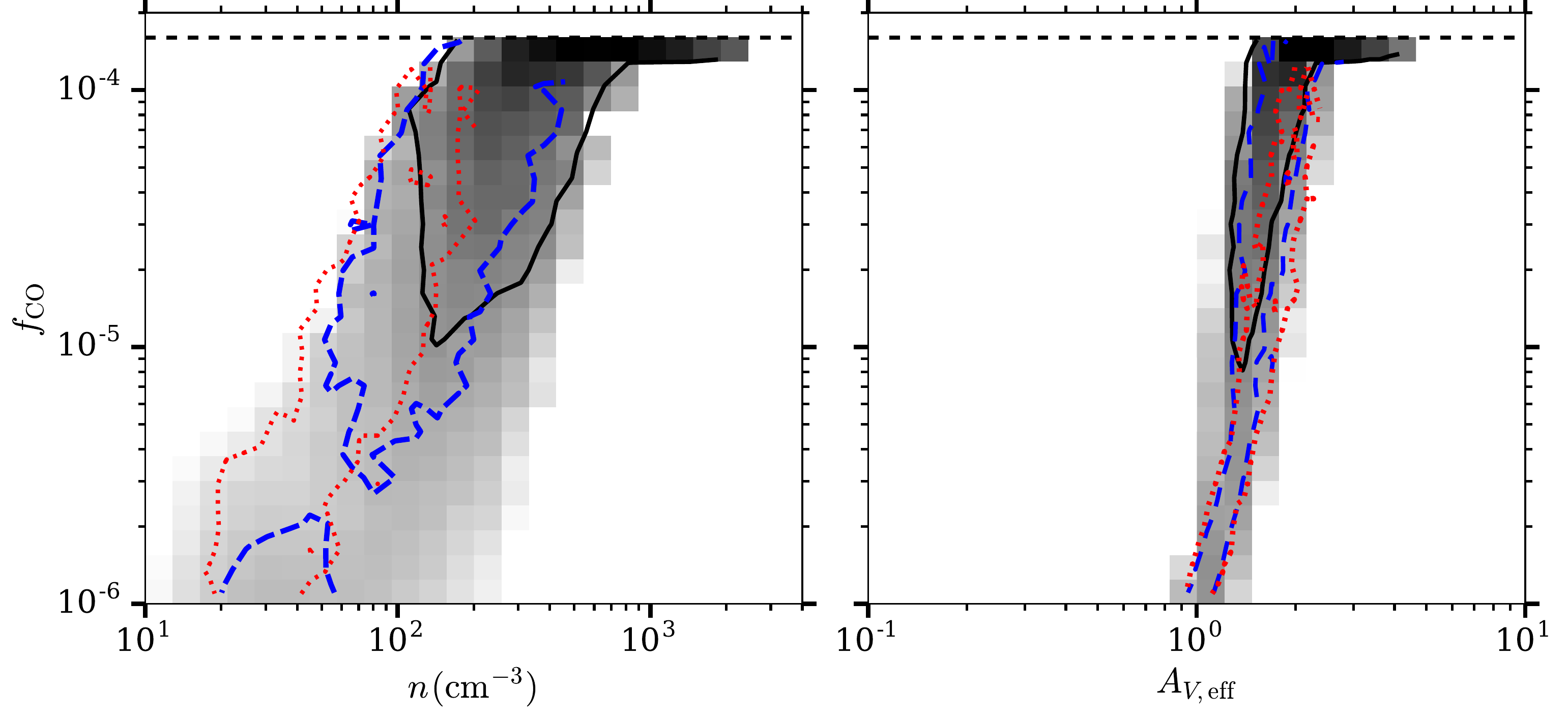}
\caption{Similar to Figure \ref{fig:fH2_n_AVsh}, but for the $\CO$ abundance
$f_\CO$.  The black dashed lines show where all carbon is in $\CO$,
i.e., $f_\CO=1.6\times 10^{-4}$.}
\label{fig:fCO_n_AVsh}
\end{figure*}

The chemical composition depends not only on density, which
affects the rate of collisional reactions, but also on shielding, which
determines the photodissociation rate by FUV photons. Which factor,
density or shielding, is more important in determining the $\Ht$ and $\CO$
abundances in realistic molecular clouds with complex structures? Figures 
\ref{fig:fH2_n_AVsh} and \ref{fig:fCO_n_AVsh} plot the
probability density distributions (PDFs) of the
$\Ht$ and $\CO$ abundances versus density and shielding in each grid cell.
We weight the PDFs by $n f_\Ht$ or $n f_\CO$, so that the color
scale is proportional to the $\Ht$ or $\CO$ mass in each bin. Simple volume
weighted PDFs will show distibutions centered at very low density and low
molecular abundances, since by volume most gas is atomic.

We quantify the shielding
by calculating the effective extinction $A_{V, \mr{eff}}$
for the photo-electric heating \citep{GOW2016},
\begin{equation}
    \chi_\mr{PE} \equiv \chi \exp(-1.8 A_{V, \mr{eff}}),
\end{equation}
where $\chi_\mr{PE}$ is the actual radiation field intensity
obtained from the six-ray radiation transfer.  

As shown in Figure \ref{fig:fH2_n_AVsh}, the $\Ht$ abundance has a much tighter
correlation with density than with shielding. This is because $\Ht$ self-shielding
is so efficient that the photodissociation rate of $\Ht$ is very
small in most regions that have a significant amount of $\Ht$. 
In the absence of photodissociation by FUV radiation, the 
$\Ht$ abundance is
then determined by the balance between $\Ht$ formation on dust grains,
\begin{equation}\label{eq:H2_gr}
    \mr{H + H + gr \rightarrow H_2 + gr},
\end{equation}
with a rate coefficient $k_\mr{gr}=3.0\times 10^{-17}~\mr{cm^3 s^{-1}}$
(assuming solar neighborhood dust abundance),
$\Ht$ formation by $\mr{H_3^+}$,
\begin{equation}\label{eq:H2_H3+}
    \mr{H_3^+ + e \rightarrow H_2 + H},
\end{equation}
with a rate coefficient $k_{\ref{eq:H2_H3+}}$,
$\Ht$ destruction by cosmic-rays,
\begin{equation}\label{eq:H2_CR}
    \mr{CR + H_2 \rightarrow H_2^+ + e},
\end{equation}
with a rate coefficient $k_\mr{CR} = 2\xi_\mr{H} (2.3 f_\Ht + 1.5 f_\mr{H})$, 
and $\Ht$ destruction by $\mr{H_2^+}$,
\begin{equation}\label{eq:H2_H2p}
    \mr{H_2^+ + H_2 \rightarrow H_3^+ + H},
\end{equation}
with a rate coefficient $k_{\ref{eq:H2_H2p}}$.
Reactions (\ref{eq:H2_CR}) and (\ref{eq:H2_H2p}) are also the main pathways for
$\mr{H_2^+}$ destruction and creation. Equilibrium of $\mr{H_2^+}$ requires
\begin{equation}\label{eq:H2p_eq}
    f_\Ht k_\mr{CR} = f_\mr{H_2^+} f_\Ht n k_{\ref{eq:H2_H2p}}.
\end{equation}
$\mr{H_3^+}$ is mainly created by reaction (\ref{eq:H2_H2p}), and destroyed by
reaction $\mr{H_3^+ + e}$, which forms $\mr{H_2 + H}$ (reaction
(\ref{eq:H2_H3+})) or $\mr{3H}$ with a branching ratio of 0.35:0.65.
Equilibrium of $\mr{H_3^+}$ requires 
\begin{equation}\label{eq:H3+_eq}
    f_\mr{H_2^+} f_\Ht n k_{\ref{eq:H2_H2p}} = \frac{1}{0.35} f_\mr{H_3^+}
    f_\mr{e} n k_{\ref{eq:H2_H3+}}.
\end{equation}
Finally, equilibrium of fully-shielded $\Ht$ 
(Equations (\ref{eq:H2_gr}) - (\ref{eq:H3+_eq})) requires
\begin{equation}\label{eq:H2_eq}
    \begin{aligned}
    f_\mr{H} n k_\mr{gr} + f_\mr{H_3^+} f_\mr{e} n k_{\ref{eq:H2_H3+}} 
    &= f_\Ht k_\mr{CR} + f_\mr{H_2^+} f_\Ht n k_{\ref{eq:H2_H2p}}\\
    f_\mr{H} n k_\mr{gr} + 0.35 f_\mr{H_2^+} f_\Ht n k_{\ref{eq:H2_H2p}}
    &= f_\Ht k_\mr{CR} + f_\mr{H_2^+} f_\Ht n k_{\ref{eq:H2_H2p}}\\
    f_\mr{H} n k_\mr{gr} &= 1.65 f_\Ht k_\mr{CR}.
    \end{aligned}
\end{equation}
In the above, each $f$ is the abundance of a given species relative to hydrogen
nuclei. 

Equation (\ref{eq:H2_eq}) can be solved with the conservation of hydrogen
nuclei 
$f_\mr{H} + 2f_\Ht = 1$, giving the equilibrium $\Ht$ abundance as a function
of $n$, plotted as the green dashed line in the left panel of Figure 
\ref{fig:fH2_n_AVsh}. This agrees very well with the upper limit of $f_\Ht$ in
the simulations. The spread of 
$f_\Ht$ at a given density is due to the incomplete shielding of FUV radiation 
in some regions where destruction of $\Ht$ from photodissociation brings its
abundance lower than that in completely shielded regions.
This can also be seen in the right panel of Figure \ref{fig:fH2_n_AVsh}: there is a
large spread of $A_{V, \mr{eff}}$ at a given $f_\Ht$, and there are many grid cells
with $A_{V, \mr{eff}} < 1$ and significant $\Ht$ abundance.

Contrary to the case of $\Ht$ abundance, which is determined mostly by density, 
the CO abundance is determined by both density and shielding, 
as shown in Figure \ref{fig:fCO_n_AVsh}. $\CO$ forms mainly in regions with 
$n \gtrsim 100~\mr{cm^{-3}}$ and $A_V \gtrsim 1$. This agrees very well with
the results from 1D slab models in \citet[][see their Figures 5 and 6]{GOW2016}.
The main reason $\Ht$ and $\CO$ form under different conditions is 
that the self-shielding of $\CO$ and
cross-shielding of $\CO$ by $\Ht$ are much less
efficient than the $\Ht$ self-shielding. As a result, $\CO$ formation is limited by
photodissociation, and $\CO$ can only form in regions with 
$A_{V, \mr{eff}} \gtrsim 1$ where the FUV radiation field is sufficiently
shielded by dust. Moreover, $\CO$ formation also requires higher
densities, as $\mr{C^+}$ and $\mr{He^+}$ formed by cosmic rays destroy $\CO$
at lower densities. Figures \ref{fig:fH2_n_AVsh} and \ref{fig:fCO_n_AVsh} again 
show that the $\Ht$ mass in our simulations is converged,
but the $\CO$ mass is not, due to the lack of resolution for very high density gas 
(see also Figure \ref{fig:nH_hist_MH2_MCO} and discussion). 

\begin{figure}[htbp]
\centering
\includegraphics[width=0.9\linewidth]{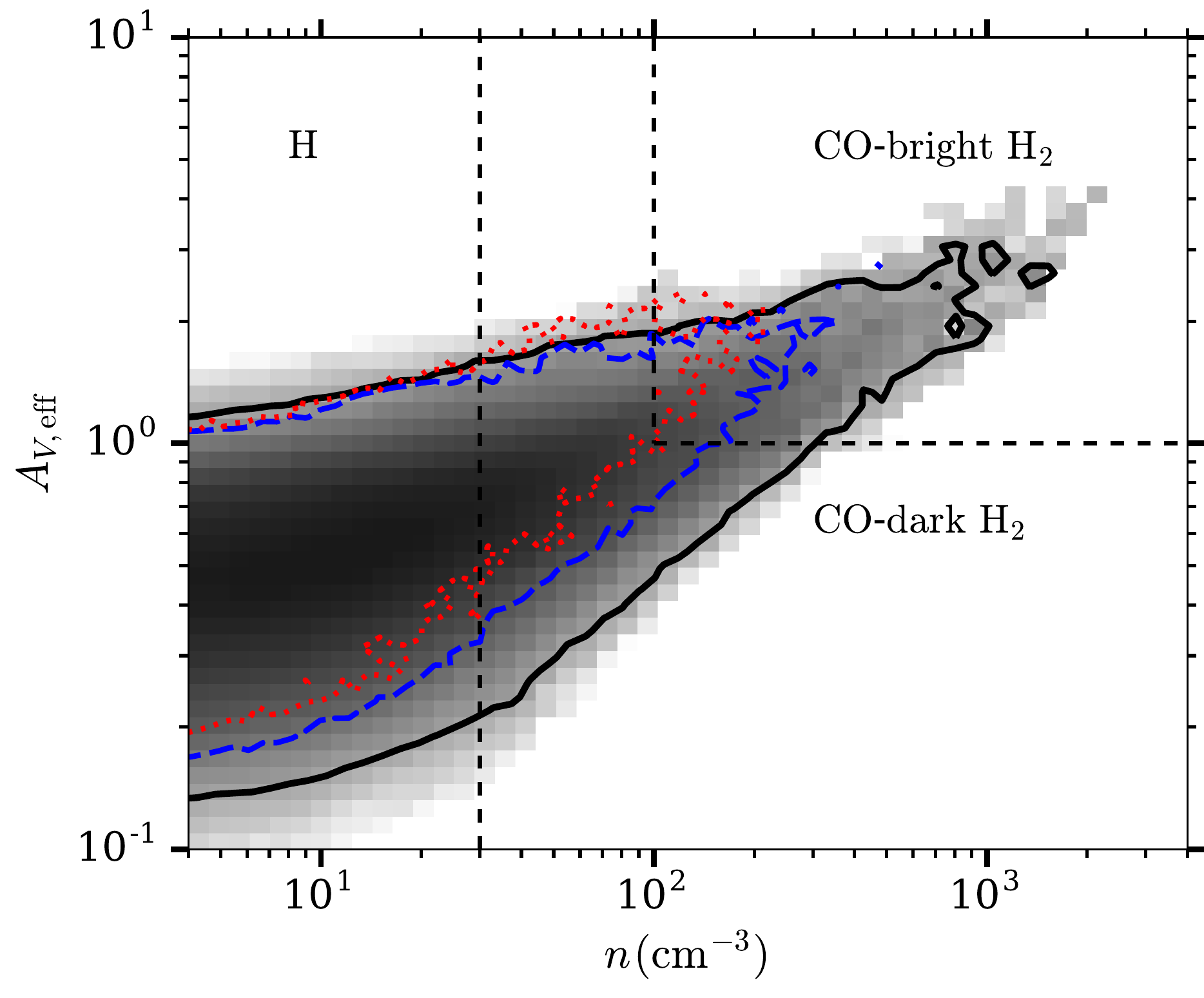}
\caption{Distribution of density $n$ versus effective extinction $A_{V,
    \mr{eff}}$. The color scale shows the log of the mass
    in each bin  in model RES-1pc, spanning across three orders of magnitude. 
    The contours indicate 99\% of the mass in models RES-1pc (black
    solid), RES-2pc (blue dashed), and RES-4pc (red dotted).
The dashed lines roughly denote the regions
where $\mr{H}$, $\Ht$, and $\CO$ form (see text in Section \ref{section:res_chemistry}).}
\label{fig:nH_AVsh}
\end{figure}

Because $\Ht$ and $\CO$ formation require different conditions, $\CO$ is only a
very approximate tracer of $\Ht$.  Figure \ref{fig:nH_AVsh} shows the distribution of 
density $n$ versus the effective extinction
$A_{V, \mr{eff}}$ for each grid cell. At a given density, there is a large range of
$A_{V, \mr{eff}}$. We roughly delineate loci where $\mr{H}$, $\Ht$, and $\CO$ form:
$\Ht$ exists in high density regions, and $f_\Ht > 0.5$ corresponds roughly to
densities $n\gtrsim 30~\mr{cm^{-3}}$. $\CO$ forms in denser
and well shielded  regions, and 
$f_\CO > 10^{-5}$ roughly corresponds to $n\gtrsim 100~\mr{cm^{-3}}$
and $A_{V, \mr{eff}} \gtrsim 1$. Figure \ref{fig:nH_AVsh} clearly
shows that a significant fraction of $\Ht$ would not be traced by $\CO$ emission 
(see $f_\mr{dark}$ in Table \ref{table:res_comparison}). As $\Delta x$
decreases from $4~\mr{pc}$ to $1~\mr{pc}$, more and more high density gas is
resolved, as also shown in Figure \ref{fig:nH_hist}. Nevertheless, for all
resolutions considered in our models, there is gas in the three different
regimes -- atomic, $\CO$-bright molecular, and $\CO$-dark molecular.

\begin{figure}[htbp]
\centering
\includegraphics[width=0.9\linewidth]{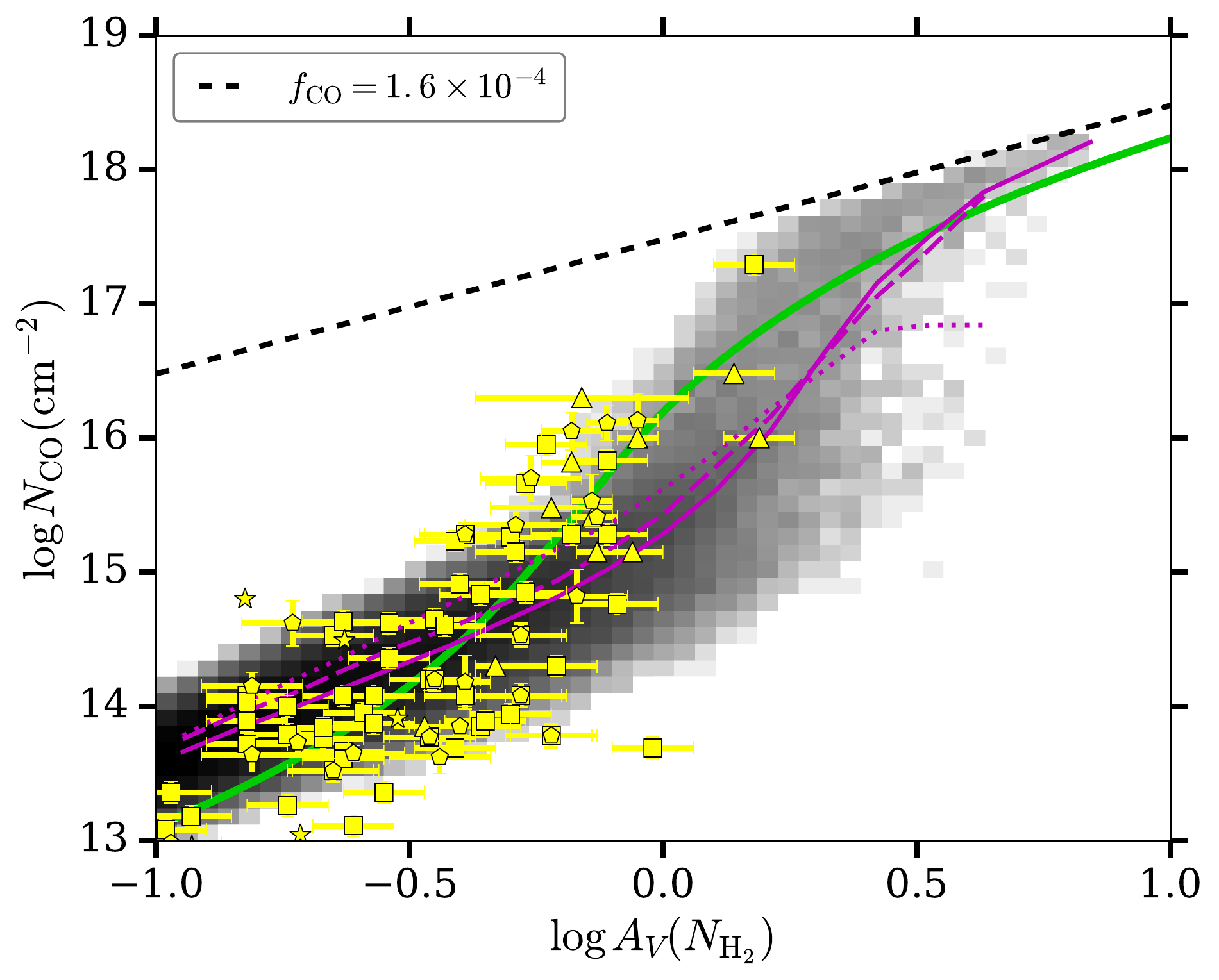}
\caption{Distribution of $\CO$ column density $N_\CO$ versus
    $A_V (N_\Ht)$ in the model RES-1pc. The color scale shows the log of the
    gas column in each bin, spanning across three orders of magnitude. The
    magenta lines indicate the median of the $\log N_\CO$ in $\log A_V (N_\Ht)$
bins for models RES-1pc (solid), RES-2pc (dashed) and RES-4pc (dotted).
The black dashed line shows
where all carbon is in $\CO$, i.e., $f_\CO=1.6\times 10^{-4}$. The yellow
symbols are UV absorption observations in
\citet{Rachford2002} (triangles), \citet{Sheffer2008} (squares), \citet{CF2004}
(stars) and \citet{BFJ2010} (pentagons), compiled by \citet{GOW2016}. The
green line shows the result from the one-sided slab model with constant
density $n=100~\mr{cm^{-3}}$ in \citet{GOW2016}.}
\label{fig:NCO_AV}
\end{figure}

To validate that we can accurately simulate chemistry in molecular clouds,
we compare the $\CO$ column densities $N_\CO$ in our simulations to that in the UV
absorption observations of diffuse molecular clouds. 
Figure \ref{fig:NCO_AV} shows the comparison between the simulations and
observations, as well as the result from the one-sided slab model in
\citet{GOW2016}.
The x-axis of Figure \ref{fig:NCO_AV} is the extinction from only $\Ht$:
\begin{equation}
    A_V(N_\Ht) = \frac{2N_\Ht}{1.87\times 10^{21}~\mr{cm^{-2}}}.
\end{equation}
To avoid
foreground/background contamination, we compare $N_\CO$ to 
$N_\Ht$ instead of the total column $N$.\footnote{\citet{GOW2016} discussed
that the dispersion in observations is much smaller when comparing $N_\CO$ to $N_\Ht$
instead of $N$.}
Compared to the simulations, the
one-sided slab model gives higher $\CO$ abundance at $A_V(N_\Ht)\sim 1$.
This is because the six-ray radiation transfer in the 3D simulations considers
extinction of FUV radiation from all directions along the Cartesian axes, 
which is generally lower than the
extinction only along the z-axis, $A_V(N_\Ht)$ (that is, $A_\mr{V, eff}
\lesssim A_V(N_\Ht)$). At $A_V(N_\Ht)\ll 1$ or 
$A_V(N_\Ht)\gg 1$, the $\CO$ abundance in the one-sided slab model and 3D simulations
are more similar, because either the FUV radiation is only weakly shielded at low
$A_V(N_\Ht)$ so that the photodissociation rate is insensitive to $A_V(N_\Ht)$,
or else already completely shielded at high $A_V(N_\Ht)$ so that the limiting factor
for $\CO$ formation is no longer photodissociation. The UV absorption
observations can only be conducted in diffuse molecular clouds with $A_V(N_\Ht)
\lesssim 1$, and there is a lack of observations at higher extinctions. For the
range of $A_V(N_\Ht)$ where the observational data are available, the
RES-1pc simulation successfully reproduce the observed range of $N_\CO$. 
Lower resolution simulations RES-2pc and RES-4pc also show similar average
values (magenta lines) and range (not shown in the Figure) of $N_\CO$ at 
$A_V \lesssim 1$. At $A_V > 1$, models with
lower resolutions start to show that the $\CO$ mass is not resolved at high
densities.

\subsubsection{Dependence of $X_\CO$ on Numerical Resolution\label{section:res_XCO}}
To understand the relation between physical properties of molecular clouds and
$\CO$ emission, a helpful reference point is 
the simple uniform slab model for molecular
clouds. In a uniform slab with constant $\CO$ excitation 
temperature $T_\mr{exc}$, Equation (\ref{eq:dInu}) can be integrated, giving
\begin{equation}\label{eq:Inu}
    I_\nu = I_\nu(0) \mr{e}^{-\tau_\nu} +
    B_\nu(T_\mr{exc})(1-\mr{e}^{-\tau_\nu}),
\end{equation}
where $S_\nu=B_\nu(T_\mr{exc})$, the blackbody
radiation field intensity at temperature $T_\mr{exc}$, and $I_\nu(0)$ is the
initial impinging radiation field intensity at $\tau_\nu=0$. 
\footnote{In observations, the intensity is often referred to as the value
after background subtraction $I_\mr{obs} = I_\nu - I_\nu(0)$. Then Equation
(\ref{eq:Inu}) is often written as 
$I_\mr{obs} = (B_\nu(T_\mr{exc}) - I_\nu(0))(1-\mr{e}^{-\tau_\nu})$.}
The line intensity $I_\nu$ is usually measured in terms of the antenna
temperature (also often referred to as the radiation temperature) in radio astronomy:
\begin{equation}\label{eq:T_A}
    T_A(\nu) = \frac{c^2}{2k\nu^2}I_\nu.
\end{equation}
In the limit of $\tau_\nu \rightarrow \infty$,
Equations (\ref{eq:Inu}) and (\ref{eq:T_A}) becomes 
\begin{equation}\label{eq:T_A_tauinf}
    T_A = \frac{T_0}{\mr{e}^{T_0/T_\mr{exc}} - 1},
\end{equation}
where $T_0=5.5~\mr{K} = h\nu_0/k$, with $\nu_0=115.3~\mr{Hz}$, the frequency of the 
$\CO (J=1-0)$ line.

Typically, the $\CO (J=1-0)$ line profile (in terms of $T_A$ and $v$)
is not too far from a Gaussian profile, and to first order,
the total $\CO$ line intensity $W_\CO$ is determined by two parameters: the
peak of the line profile $T_\mr{peak}$ and the width/velocity
dispersion of the line $\sigma_v$. Under the assumption that the line center is
optically thick so that Equation (\ref{eq:T_A_tauinf}) applies, the observed
peak antenna temperature, $T_\mr{peak}$, would be directly related to the
excitation temperature $T_\mr{line}$,
\begin{equation}\label{eq:Tline}
    T_\mr{line} \equiv \frac{5.5~\mr{K}}{\ln (5.5~\mr{K}/T_\mr{peak}+1)}.
\end{equation}
We use the notation $T_\mr{line}$ for the excitation temperature 
derived from the line profile ($T_\mr{peak}$) to distinguish 
from the true excitation temperature in the molecular clouds $T_\mr{exc}$. 
Although $T_\mr{line}=T_\mr{exc}$ in a uniform slab cloud as long as
the $\CO$ line center is optically thick, in real molecular clouds and also in
our numerical simulations, the excitation
temperature along the line of sight is not constant, and $T_\mr{line}$ serves as
an estimate of the excitation temperature where most $\CO$ emission comes
from. For $T_\mr{peak}\gtrsim 5.5~\mr{K}$, Equation (\ref{eq:Tline}) gives 
$T_\mr{line} \approx T_\mr{peak}$. Another important parameter for the $\CO$
line, the velocity dispersion, is calculated using
$\sigma_v\equiv \sqrt{\langle v^2 \rangle_{T_A} - \langle v \rangle_{T_A}^2}$,
where $\langle v \rangle_{T_A}\equiv \int v T_A\di v/\int T_A \di v$ 
is the intensity weighted average of velocity, and similarly 
$\langle v^2 \rangle_{T_A}\equiv \int v^2 T_A\di v/\int T_A \di v$. 

\begin{figure*}[htbp]
\centering
\includegraphics[width=0.9\linewidth]{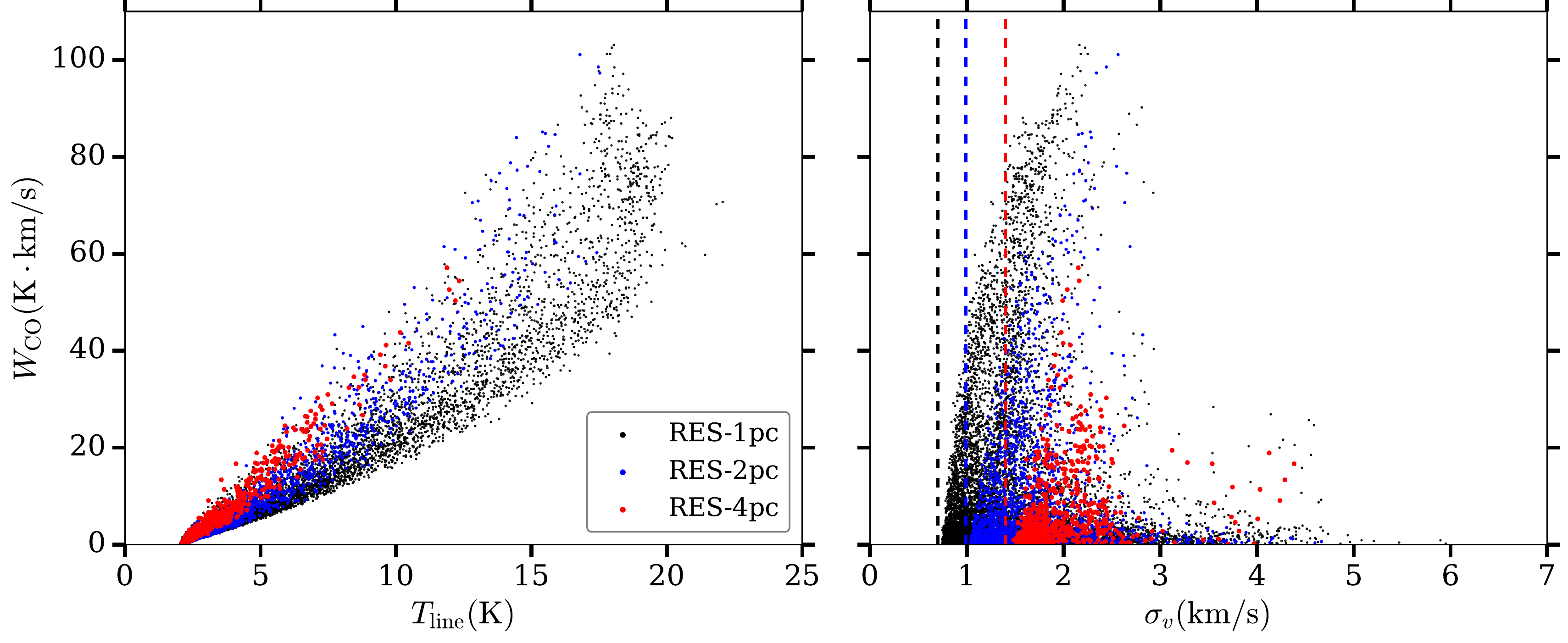}
\caption{{\sl Left}: scatter plot of $W_\CO$ vs. $T_\mr{line}$, 
    the excitation temperature of the $\CO(J=1-0)$ at 
    line center (see Equation (\ref{eq:Tline})); and {\sl right}: $W_\CO$ vs.
    the velocity dispersion of the line, $\sigma_v$. Both panels show
    models RES-1pc (black), RES-2pc (blue) and RES-4pc (red), with
    the area of points proportional
to the area of the pixel at the corresponding resolution. The vertical dashed
lines show the sub-grid micro-turbulence parameter (see Equation (\ref{eq:v_turb})).}
\label{fig:Texc_sigmav_WCO}
\end{figure*}

The relations between $W_\CO$ and $T_\mr{line}$ or $\sigma_v$ in models RES-1pc,
RES-2pc and RES-4pc are shown in Figure \ref{fig:Texc_sigmav_WCO}.
$T_\mr{line}$ ranges from $\sim 2~\mr{K}$ (from the CMB background)
to $\sim 20~\mr{K}$ (from the kinetic temperature of dense gas as discussed below),
similar to the range of excitation temperature
observed in Perseus and Taurus molecular clouds \citep{Pineda2008, Pineda2010}.
The velocity dispersion spans a relatively narrow range $\sigma_v \approx
1-2~\mr{km/s}$, and the lower limit for $\sigma_v$ is set by the sub-grid
micro-turbulence velocity in Equation (\ref{eq:v_turb}). The
observations of nearby molecular clouds have higher resolutions of 
$\sim 0.2-0.4~\pc$, and therefore a slightly lower but still limited range of
velocity dispersions 
$\sigma_v \approx 0.8-1.5~\mr{km/s}$ \citep{Pineda2010, Kong2015}.  
$W_\CO$ increases with both $T_\mr{line}$ and $\sigma_v$. For a Gaussian
profile with $T_\mr{peak}\gtrsim 5.5~\mr{K}$, $W_\CO = \sqrt{2\pi}T_\mr{peak}\sigma_v
\approx \sqrt{2\pi}T_\mr{line}\sigma_v$. Because the variation in $\sigma_v$ is small, 
$W_\CO$ correlates very well with $T_\mr{line}$, except for regions where $T_\mr{line}$
saturates around $20~\mr{K}$. There is no saturation
of $W_\CO$, and $W_\CO$ keeps increasing with increasing $\sigma_v$.  

\begin{figure}[htbp]
\centering
\includegraphics[width=0.9\linewidth]{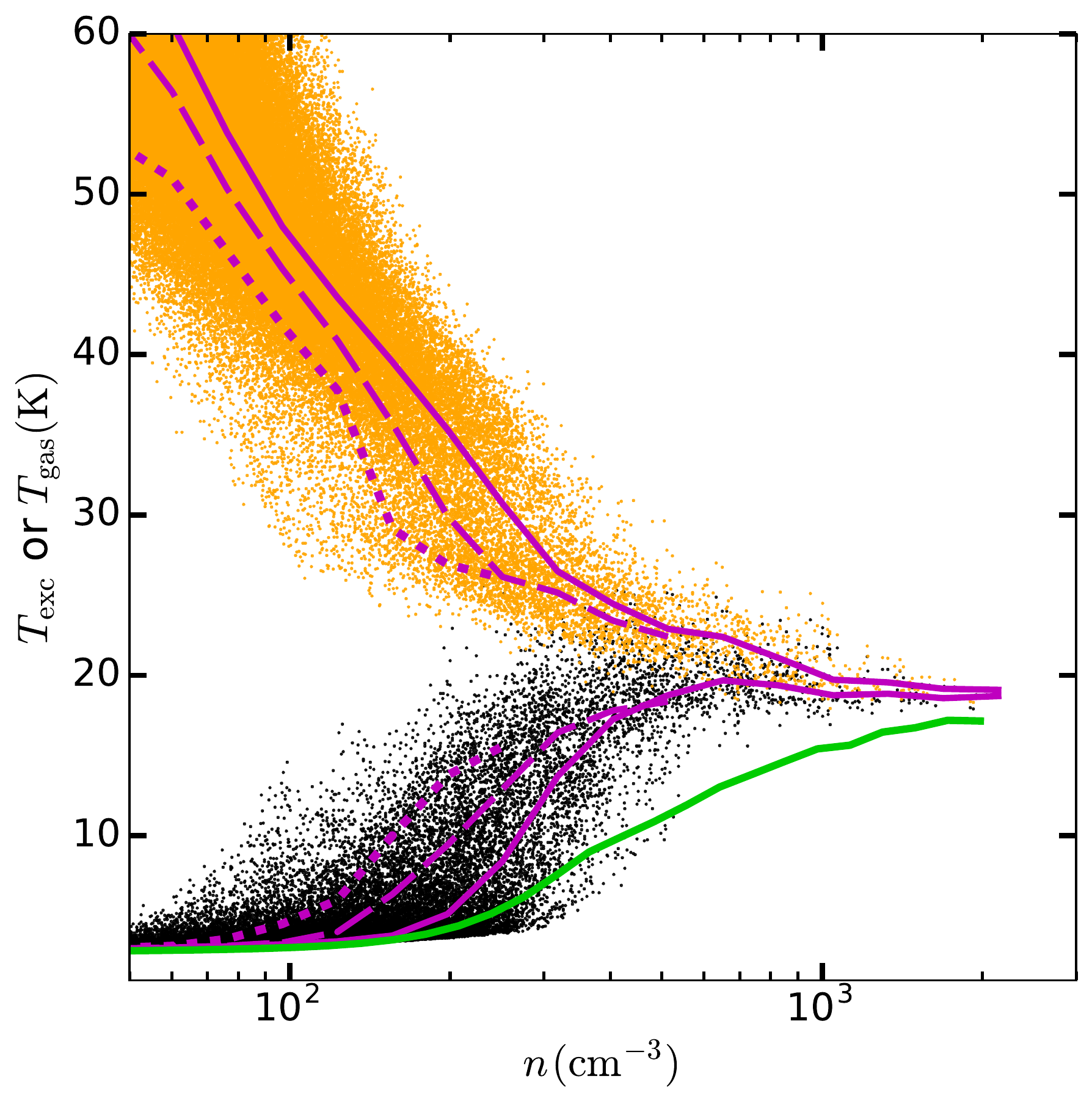}
\caption{Scatter plot of the gas temperature 
        (orange, upper branch) and excitation
    temperature of $\CO(J=1-0)$ line (black, lower branch) 
versus gas density $n$ in each cell for model RES-1pc. The magenta lines
indicate the median gas temperature and $\CO$ excitation temperature in density
bins for moedels RES-1pc (solid), RES-2pc (dashed), and RES-4pc (dotted).
The green line
is the estimation of $T_\mr{exc}$ in a two-level system model
(see Section \ref{section:res_XCO}). }
\label{fig:Texc_nH}
\end{figure}

$W_\CO$ is largely determined by the excitation temperature, and the excitation
temperature in turn depends on the gas density and temperature.
Figure \ref{fig:Texc_nH} shows the excitation temperature $T_\mr{exc}$ and gas
temperature $T_\mr{gas}$ versus the gas density in each grid cell. 
$T_\mr{gas}$ decreases with increasing density, as the gas cooling
becomes more efficient, and heating is also reduced by shielding of the FUV radiation
field in dense regions. On the other hand, 
$T_\mr{exc}$ increases with increasing density, because the collisional
excitation rate of $\CO$ is proportional to density, 
and because radiative trapping increases in dense regions. 

The lower solid magenta line in Figure \ref{fig:Texc_nH} shows the median
$T_\mr{exc}$ from model RES-1pc as a function of density.  
$T_\mr{exc}$ only reaches equilibrium with $T_\mr{gas}$ at 
$n \gtrsim 400~\mr{cm^{-3}}$, implying that local thermal equilibrium (LTE)
approximation would fail in most regions.

At a given density, $T_\mr{exc}$ is higher at lower resolutions for two reasons.
First, the velocity gradient $|\di v/\di r|$ is smaller at lower resolutions,
leading to higher $\tau_\mr{LVG}$ and thus lower escape probability $\beta$ 
and higher $T_\mr{exc}$ at a given density
(See Equations (\ref{eq:n12n0}) and (\ref{eq:TexcApprox}) below). Second, 
at lower resolutions, 
less high-density gas is resolved, and a larger fraction of the $\CO$ gas is 
found in lower-density gas (see Fig. 
  \ref{fig:nH_hist_MH2_MCO}b).  This shifts the distributions of
$T_\mr{exc}$ and $T_\mr{gas}$ in Figure \ref{fig:Texc_nH} to the left at
lower resolution in models RES-2pc and RES-4pc (dashed and
dotted magenta lines).


In general, thermalization is expected
  for densities above a critical value at which collisional deexcitation
  exceeds spontaneous emission.
For $\CO$ collisions with $\Ht$, the collisional deexcitation rate is
$n_\mr{H_2} k_{10}$ for 
\begin{equation}\label{eq:k_10}
    k_{10} \approx 6\times 10^{-11} 
    \left( \frac{T_\mr{gas}}{100~\mr{K}} \right)^{0.2}~\mr{cm^3 s^{-1}},
\end{equation}
at $10~\mr{K}\lesssim T_\mr{gas} \lesssim 250~\mr{K}$ \citep{FL1985, Flower2001,
  Draine2011}. The spontaneous emission rate is $\beta A_{10}$, where
Equation (\ref{eq:beta}) gives the escape probability $\beta$, so that 
\begin{equation}\label{eq:ncrit}
    n_\mr{crit} = \frac{\beta A_{10}}{k_{10}}.
\end{equation}
For $T_\mr{gas}=20~\mr{K}$, Equation (\ref{eq:k_10}) gives
$A_{10}/k_{10}  \approx 2.1\times 10^3 ~\mr{cm^{-3}}$.

\begin{figure}[htbp]
\centering
\includegraphics[width=0.9\linewidth]{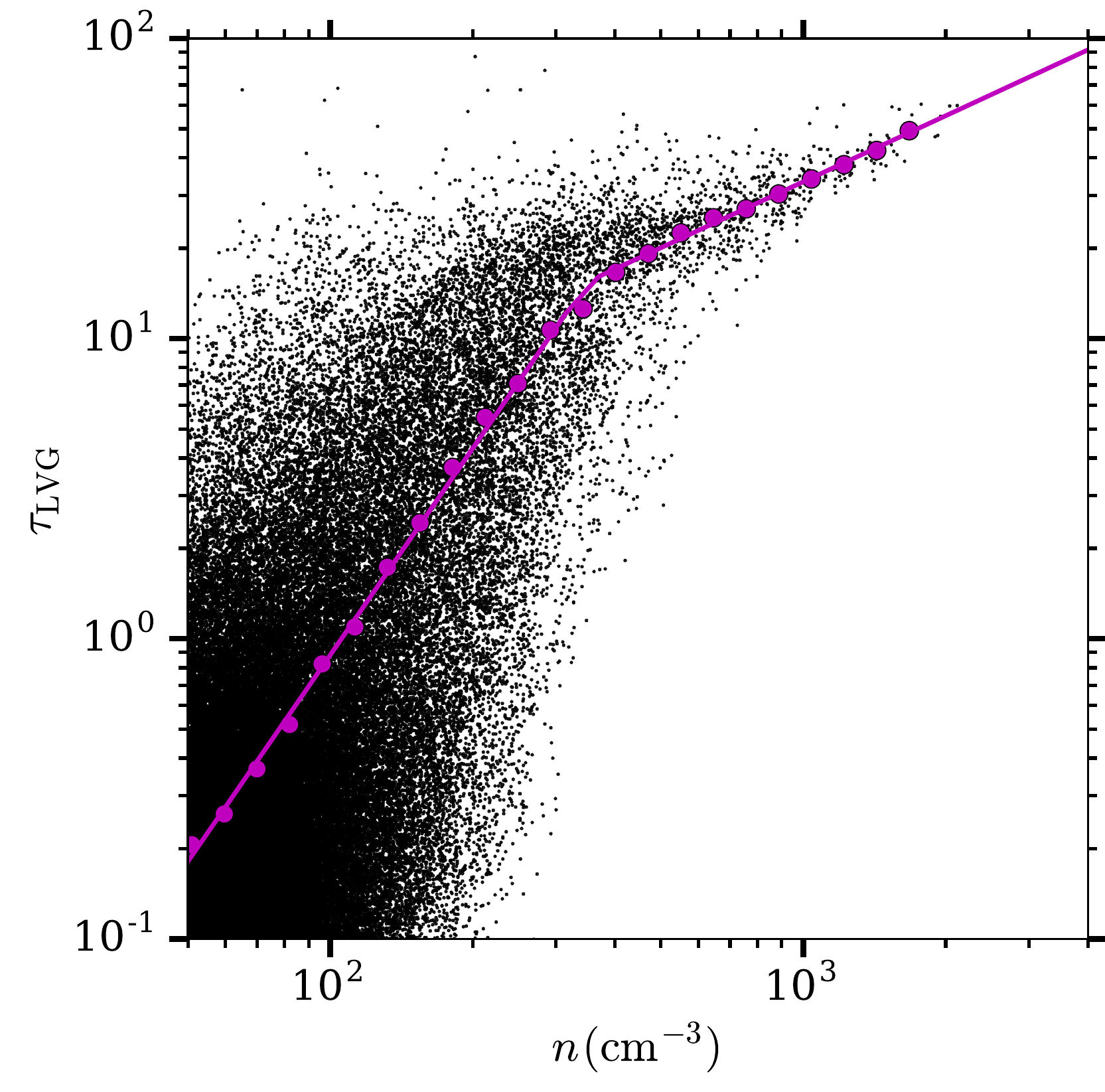}
\caption{Scatter plot of the optical depth from the LVG approximation $\tau_\mr{LVG}$
vs. density $n$ in each grid cell in model RES-1pc. The magenta circles are the
binned average of $\tau_\mr{LVG}$, and the line is a broken power-law fit to 
the circles (Equation (\ref{eq:tau_LVG_fit})). }
\label{fig:tau_LVG}
\end{figure}

With increasing density, the optical depth
$\tau_\mr{LVG}$ increases, leading to
decreasing $\beta$ (Figure \ref{fig:tau_LVG});
at large $\tau_\mr{LVG}$, $\beta \approx 1/\tau_\mr{LVG}$. 
For model RES-1pc, we fit the average
$\tau_\mr{LVG}$ at a given
density with a broken power-law (magenta line in Figure \ref{fig:tau_LVG}):
\begin{equation}\label{eq:tau_LVG_fit}
    \begin{aligned}
        \tau_\mr{LVG} &= 2.4\times 10^{-5} (n/\mr{cm^{-3} })^{2.3}, 
        \quad &n < 350~\mr{cm^{-3}}\\
        \tau_\mr{LVG} &= 0.21 (n/\mr{cm^{-3} })^{0.73}, \quad &n \geq 350~\mr{cm^{-3}}.
    \end{aligned}
\end{equation} 

Combining Equations~(\ref{eq:ncrit}) and (\ref{eq:tau_LVG_fit}) yields
$n_\mr{crit} \sim 300 ~\mr{cm^{-3}}$.  
Thus, in regions where $n \gtrsim 400~\mr{cm^{-3}}$,
the $\CO(J=1)$ level is expected to be thermalized, and this is indeed
consistent with the median $T_\mr{exc}$ for model RES-1pc.\footnote{
We note that $\tau_\mr{LVG}$ depends on the density and velocity structure,
which is resolution dependent, so the density for thermalization
is not expected to be the same for models RES-2pc and RES-4pc as for model
RES-1pc. In fact, the velocity gradient $|\di v/\di r|$ 
is smaller at lower resolutions, leading to higher average $\tau_\mr{LVG}$,
and lower density for thermalization in
models RES-2pc and RES-4pc (see Equation (\ref{eq:tau_LVG}) and discussions of
Figure \ref{fig:Texc_nH}).
}

The dependence of $T_\mr{exc}$ on $n$ can be understood in a
simplified 2-level system model. The excitation temperature is defined as
\begin{equation}\label{eq:T_exc}
T_\mr{exc} \equiv \frac{T_0}{\ln \left(\frac{n_0/g_0}{n_1/g_1}\right)  }. 
\end{equation}
With the escape probability approximation, the level populations are given by
\citep{Draine2011} 
\begin{equation}\label{eq:n12n0}
    \frac{n_1}{n_0} = \frac{n_c k_{01} + \frac{g_1}{g_0}\beta A_{10}
    n_\gamma^{(0)}}{n_c k_{10} + \beta A_{10}(1 + n_\gamma^{(0)})},
\end{equation}
where
\begin{equation}
    k_{01} = \frac{g_1}{g_0}k_{10}\mr{e}^{-T_0/T_\mr{gas} },
\end{equation}
$n_c$ is the number density for the collisional species, 
and
$n_\gamma^{(0)} = 1/(\mr{e}^{T_0/T_\mr{CMB}}-1)$ the background
incident radiation field from the CMB.
If the CMB terms are negligible,
Equation (\ref{eq:T_exc}) becomes
\begin{equation}\label{eq:TexcApprox}
T_\mr{exc} = \frac{T_\mr{gas}}{1+ \frac{T_\mr{gas}}{T_0}\ln (1 +
  \frac{\beta A_{10}}{n_c k_{10}})}.
\end{equation}
For $\beta/n_c$ small,
$T_\mr{exc}\rightarrow T_\mr{gas}$.

The excitation temperature can be estimated as a function of density by
Equations (\ref{eq:beta}), (\ref{eq:T_exc}), (\ref{eq:n12n0}) and
(\ref{eq:tau_LVG_fit}) (assuming $\tau=\tau_\mr{LVG}$ in Equation
(\ref{eq:beta}) and using the average value of $T_\mr{gas}$ at a given density).
The analytic 2-level system 
approximation for simulation RES-1pc (green line)
agrees well  with the result from radiation transfer
by the RADMC-3D code (lower solid magenta line)
at low and high densities,
while there are differences within a factor of two at intermediate densities $n\sim
300$--$1000~\mr{cm^{-3}}$.
This is because the $\CO$ rotational levels $J=1$, $2$, and $3$ 
have energies of $5.5$, $16.6$, and $33.2~\mr{K}$, all lower or comparable to
the gas temperature. Indeed, there are significant populations in the $J\ge 2$ levels,
as expected given that $T_\mr{gas} > 5.5~\mr{K}$. The
analytical expression in Equation (\ref{eq:n12n0}) only takes into account 
the $J=0$ and $J=1$ levels, and therefore cannot predict the
excitation temperature very accurately. At low and high densities the
differences are small because the excitation temperature there is determined by 
the background CMB temperature or the gas temperature as the $\CO$ rotational
levels approach LTE.  Nonetheless, the analytical 2-level
system approximation agrees with the general
trend from the radiation transfer calculations, 
and gives some insight into the relation between $T_\mr{exc}$, $T_\mr{gas}$ and $n$.
As a further test, we ran RADMC-3D only including 
    the first $J=0$ and $J=1$ rotational levels of $\CO$,
and found that it can indeed reproduce the analytical result of the 2-level
system model. Figure \ref{fig:Texc_nH_2level} shows this comparison.

\begin{figure*}[htbp]
\centering
\includegraphics[width=0.9\linewidth]{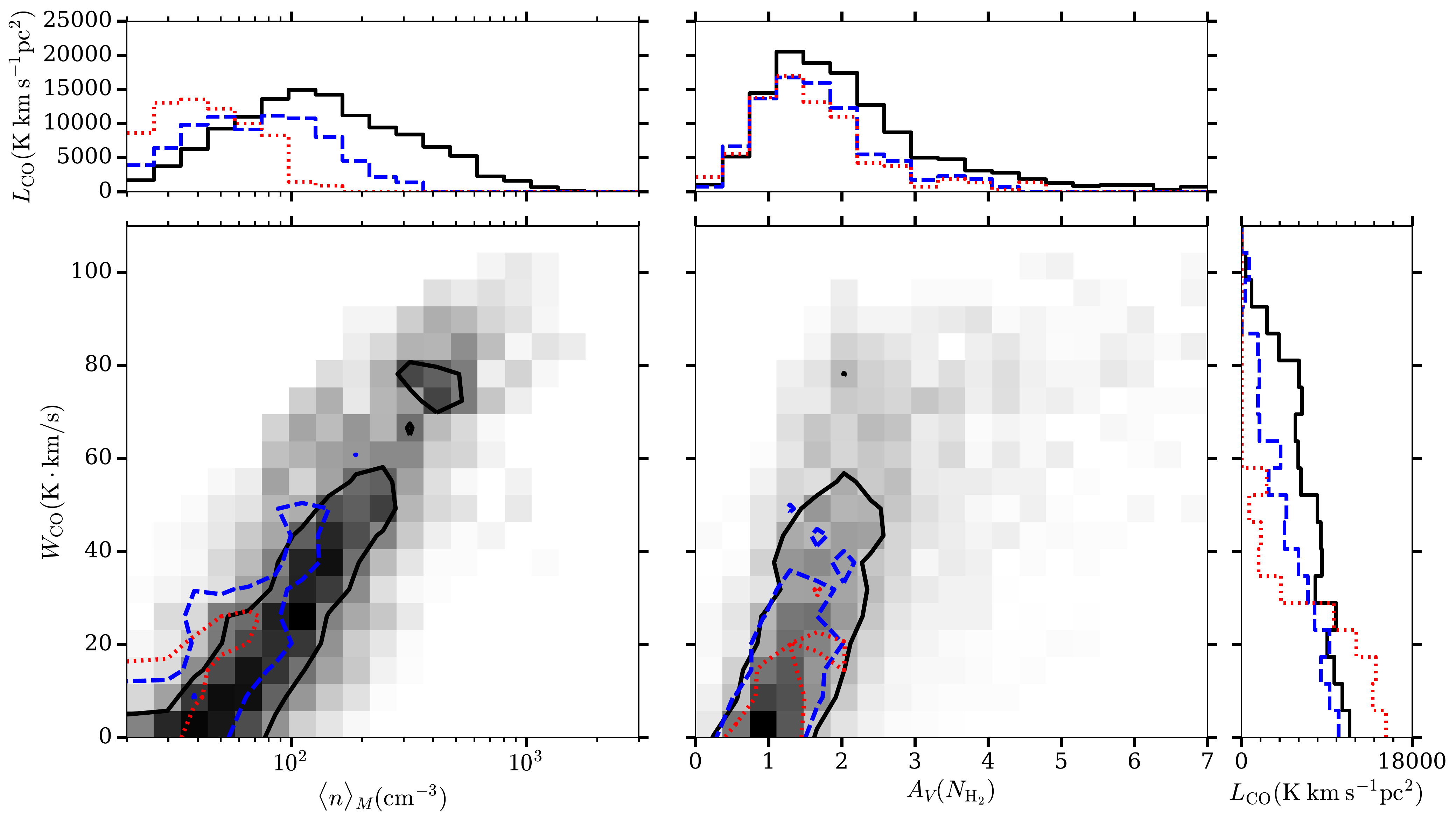}
\caption{Distributions of $W_\CO$ versus  mass-weighted mean density 
    $\langle n \rangle_M$ ({\sl left}) and $A_V(N_\Ht)$ ({\sl right}).
    The color scale in the PDF is proportional to
    the total $\CO$ luminosity $L_\CO$ from each bin in model RES-1pc. 
    The black solid contour of the color scale indicates the regions where 50\% of
    the total $\CO$ emission comes from in model RES-1pc. The blue dashed and red dotted 
    contours indicate 50\% of the emission,
    but for models RES-2pc and RES-4pc. The 1D histograms show
    $L_\CO$ in each bin for models RES-1pc (black solid), RES-2pc (blue dashed)
    and RES-4pc (red dotted).
}
\label{fig:WCO_nHM_AVH2_hist}
\end{figure*}

The relation between $W_\CO$ and $T_\mr{line}$, as well as the
relation between $T_\mr{exc}$ and density, give rise to the strong correlation between
$W_\CO$ and the average (mass weighted) density $\langle n \rangle_M$
along the line of sight (Figure \ref{fig:WCO_nHM_AVH2_hist} left panel).
Moreover, we found that in the simulations, $N_\Ht$ increases
systematically with $\langle n \rangle_M$ (see Figure
\ref{fig:nH_AVsh}).
This results in a correlation between $W_\mr{CO}$ and $N_\Ht$ 
(Figure \ref{fig:WCO_nHM_AVH2_hist} right panel). Although $X_\CO$ is measured
in terms of $W_\CO$ and $N_\Ht$, there is a smaller dispersion in the
correlation between $W_\CO$ and $\langle n \rangle_M$. This suggests that the 
$\CO$ emission is more fundamentally a measure of $\Ht$ density than column
density.

\begin{figure}[htbp]
\centering
\includegraphics[width=0.9\linewidth]{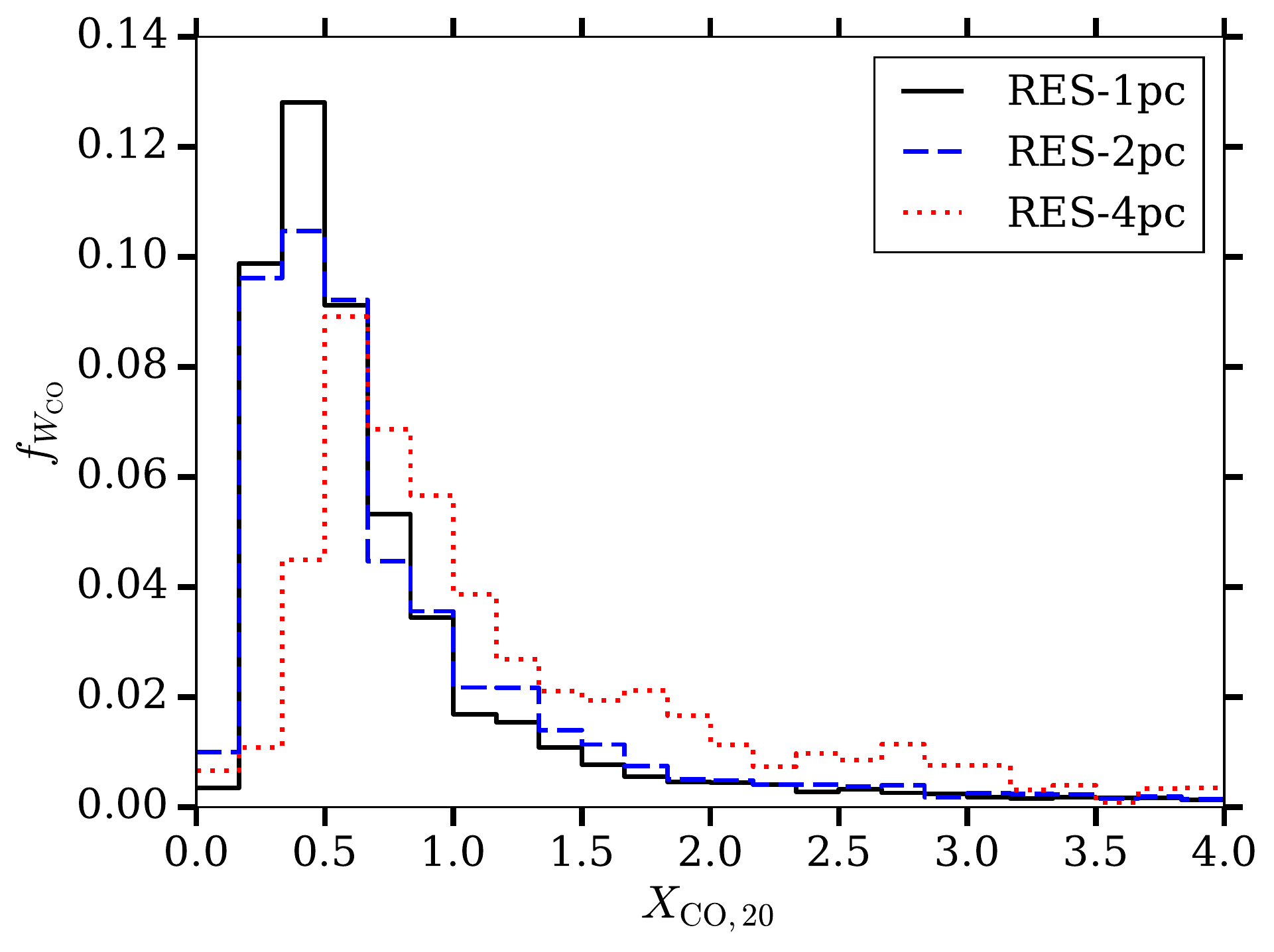}
\caption{Histograms of $X_\mr{CO, 20}$ weighted by $W_\CO$, 
    in models RES-1pc (black solid), RES-2pc (blue dashed) and RES-4pc (red dotted).
}
\label{fig:XCO_hist}
\end{figure}

The effect of numerical resolution on $W_\CO$ is already evident in Figures
\ref{fig:Texc_nH} and \ref{fig:WCO_nHM_AVH2_hist}. As the resolution increases,
more high density gas forms in the simulation, and thus there are
more pixels with high
$W_\CO$. Numerical resolution also has an effect on $X_\CO$, as shown in 
Figure \ref{fig:XCO_hist}, the histogram of
$X_\mr{CO,20}=X_\mr{CO}/(10^{20}~\mr{cm^{-2}K^{-1}km^{-1}s})$ weighted by
$W_\mr{CO}$. The average $X_\CO$ in a certain region can be written as:
\begin{equation}
    \langle X_\CO \rangle = \frac{\sum N_\Ht}{\sum W_\CO}
    = \frac{\sum \frac{N_\Ht}{W_\CO} W_\CO }{\sum W_\CO}
    = \frac{\sum X_\CO W_\CO}{\sum W_\CO}.
\end{equation}
In other words, $\langle X_\CO \rangle$ is simply the $W_\CO$
weighted average of $X_\CO$ in each pixel. Therefore, the peak of the histogram in
Figure \ref{fig:XCO_hist} roughly indicates the average $X_\CO$ in the whole
simulation domain. The distributions of $X_\CO$ in models RES-1pc and RES-2pc
are very similar, with a slightly higher peak in RES-1pc. As a result, 
$\langle X_\CO \rangle$ is almost the same in RES-1pc and RES-2pc (Table
\ref{table:res_comparison}). The model RES-4pc,
however, peaks at larger $X_\CO$ than the higher resolution models, and therefore
has a higher $\langle X_\CO \rangle$. This is because the peak of $X_\CO$
distribution, $X_\mr{CO,20}\approx 0.5$, comes from regions with moderatly high
density $n\approx 100~\mr{cm^{-3}}$ and $\CO$ emission
$W_\CO \approx 40~\mr{K\cdot km/s}$, which can only be resolved at a
resolution finer than $2~\pc$ (see histograms of $\langle n \rangle_M$
and $W_\CO$ in Figure \ref{fig:WCO_nHM_AVH2_hist}). 
Therefore, we conclude that a numerical resolution of at least $2~\pc$ is needed
in order to resolve the average $X_\CO$ in molecular clouds for solar
neighborhood conditions.

\begin{figure*}[htbp]
\centering
\includegraphics[width=0.9\linewidth]{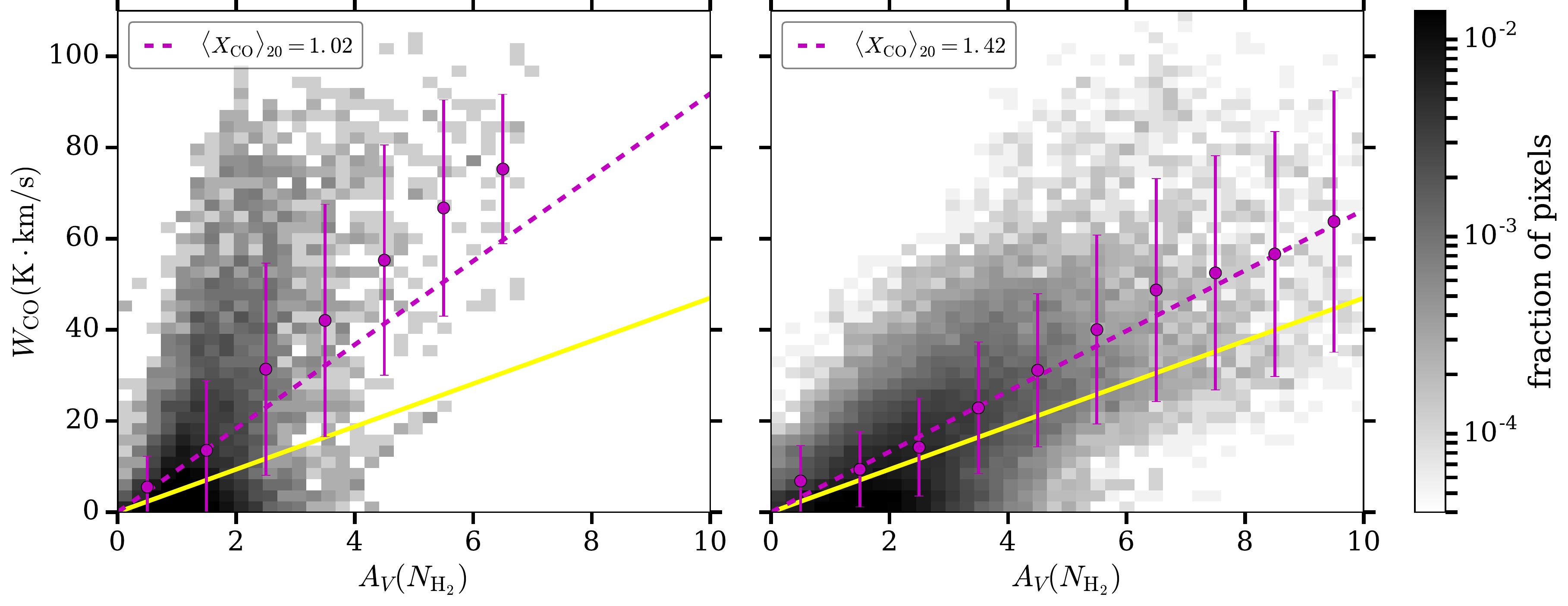}
\caption{{\sl Left}: Distribution of $W_\CO$ versus $A_V(N_\Ht)$ in model RES-1pc.
    The magenta filled circles and error bars show the average value and
    standard deviation of $W_\CO$ in each $A_V(N_\Ht)$ bin.
    The yellow solid line shows the standard Milky Way value of $X_\mr{CO,MW,20}=2$. 
    The magenta dashed line shows the average $X_\CO$ in model RES-1pc,
    $\langle X_\mr{CO} \rangle_{20}=1.02$. {\sl Right}: similar to the left
    panel, but for the observations of Orion A and B molecular clouds by
    \citet{Ripple2013}. The average $X_\CO$ for the Orion A and B clouds 
    is $\langle X_\mr{CO} \rangle_{20}=1.42$.}
\label{fig:WCO_AVH2}
\end{figure*}

Finally, we compare the distribution of $W_\CO$ versus $A_V(N_\Ht)$ in model RES-1pc
to observations of the Orion A and B molecular clouds by \citet{Ripple2013}, as
shown in Figure \ref{fig:WCO_AVH2}. Considering the noise level in
the observation, we use a higher 
threshold of $W_\CO>1~\mr{K\cdot km/s}$ to compare to the $\CO$-bright region in Orion.
Because most $\CO$ emission comes from regions with
$W_\CO \gg 1~\mr{K\cdot km/s}$, $X_\CO$ is not sensitive to the $W_\CO$
threshold.
Our simulations shows a similar distribution of $W_\CO$ versus $A_V(N_\Ht)$ to
that in Orion.  The dispersion of $W_\CO$ at 
a given $A_V(N_\Ht)$ is large, as much as more than an order of magnitude at
low $A_V(N_\Ht)$. However, despite the large dispersion
of $W_\CO$, the average $X_\CO$ (which is inversely proportional to the slope)
in different $A_V(N_\Ht)$ bins is very similar, only varying by a factor of
$\sim 2$. Similar features are observed in many Milky Way molecular clouds by
\citet{Lee2018}, and we discuss this in more detail in Section
  \ref{section:result:2pc}.

There are also some differences between the simulation and observations. 
The average $X_\CO$ in RES-1pc is a factor of 1.4 lower than that in Orion.
As we shall show based on other analyses and comparisons, the 
typical $X_\CO$ in our simulations is about a factor of $\sim 2$ lower than the
standard Milky Way value;
we discuss possible reasons 
for this discrepency at the end of Section \ref{section:result:2pc}.
We also note that 
because the observation in \citet{Ripple2013} has a higher
spatial resolution of $\sim 0.2~\pc$, there are more pixels at $A_V(N_\Ht)
\gtrsim 4$ in the observation in Figure \ref{fig:WCO_AVH2}.
The simulation has more pixels at 
$W_\CO \gtrsim 60~\mr{K\cdot km/s}$, a result of the slightly higher velocity
dispersions (see Figure \ref{fig:Texc_sigmav_WCO} and discussion). 
In spite of these differences, the
general good agreement between the models and observations indicates
that the simulations can succesfully reproduce the basic physical properties of
observed molecular clouds. 

\subsection{Non-equilibrium Chemistry\label{section:non-equilibrium}}
The realistic ISM is highly dynamical: turbulence constantly
creates and disperses molecular clouds, and moves gas to environments with
different density, temperature and radiation field strength.
As a result, non-equilibrium
chemistry is likely to be important, especially in low density diffuse gas
where the chemical timescales are long compared to the dynamical timescales. This
is especially an issue for $\Ht$.  Molecular hydrogen
can form in low density gas due
to its effective self-shielding, but its formation timescale, 
$t_\Ht \approx 10 \Myr~(n/100~\mr{cm^{-3}})^{-1}$ \citep{GOW2016}, can be
longer than the dynamical timescale (Equation (\ref{eq:t_turb})). Because
$\CO$ formation chemically relies on the existence of $\Ht$, the $\CO$ abundance
in molecular clouds can also be far from equilibrium. 
In this section, we carry out comparisons between models with
different $t_\mr{chem}$ (model IDs start with TCHEM in Table \ref{table:model})
to investigate the effect of non-equilibrium chemistry on $X_\CO$.

\begin{figure}[htbp]
\centering
\includegraphics[width=0.9\linewidth]{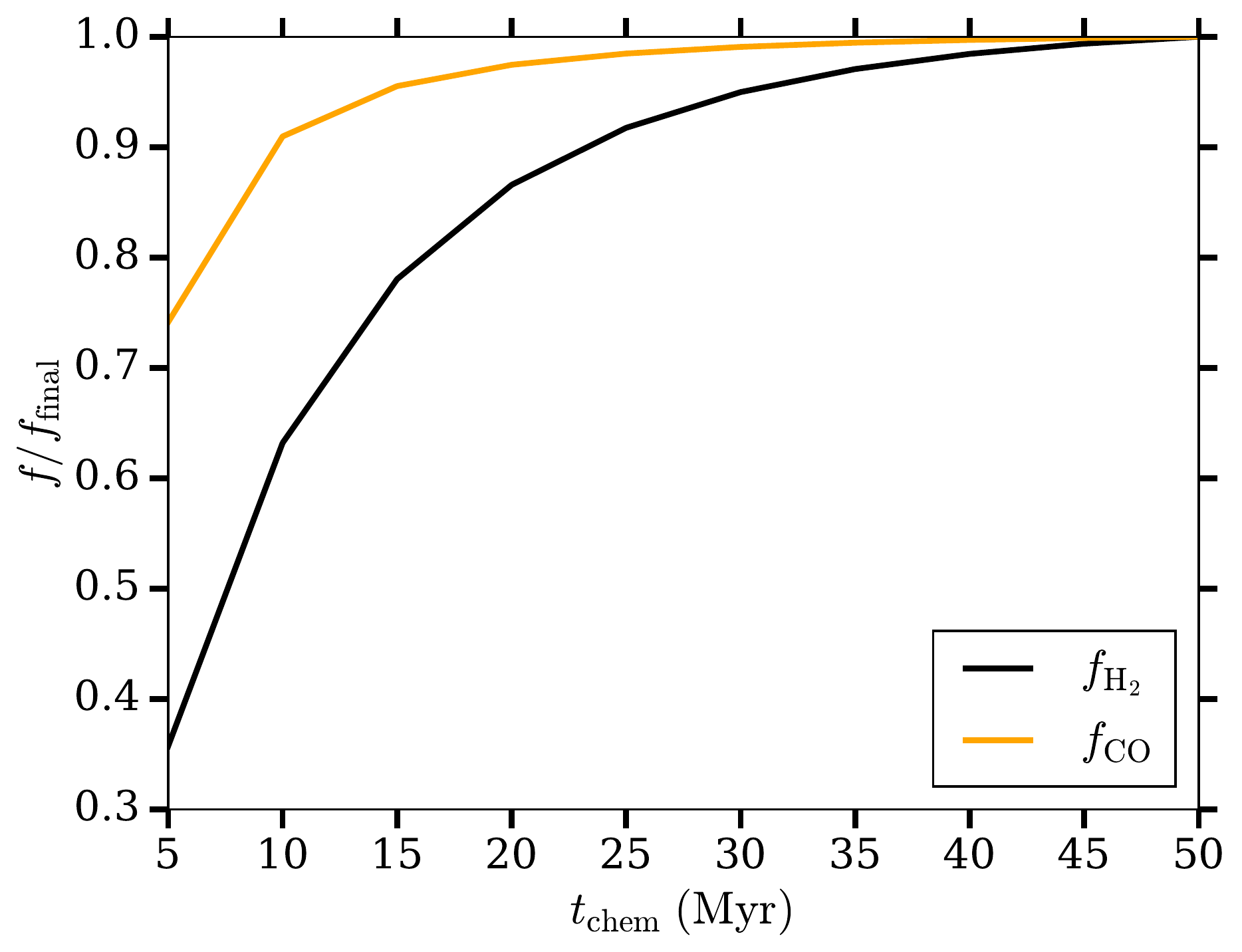}
\caption{The abundance of $\Ht$ (black) and $\CO$ (orange)
    as a function of $t_\mr{chem}$ in TCHEM models. The y-axis is normalized by the
    final $\Ht$ or $\CO$ abundance at $t_\mr{chem}=50~\Myr$.
}
\label{fig:fCO_fH2_all_t}
\end{figure}

Both $\Ht$ and $\CO$ abundance increase over $t_\mr{chem}$, 
reaching a steady state at $t_\mr{chem} \approx 50~\Myr$, 
as shown in Figure \ref{fig:fCO_fH2_all_t}. Over timescales relevant for
clouds of size $\gtrsim 10~\pc$ (Equation (\ref{eq:t_turb})),
there is a larger increase in the 
$\Ht$ abundance than $\CO$: From $t_\mr{chem}=5~\Myr$ to $50~\Myr$, the $\Ht$
abundance increases by a factor of $\sim 3$, while $\CO$ abundance increases
only by $\sim 30\%$.

\begin{figure}[htbp]
\centering
\includegraphics[width=0.9\linewidth]{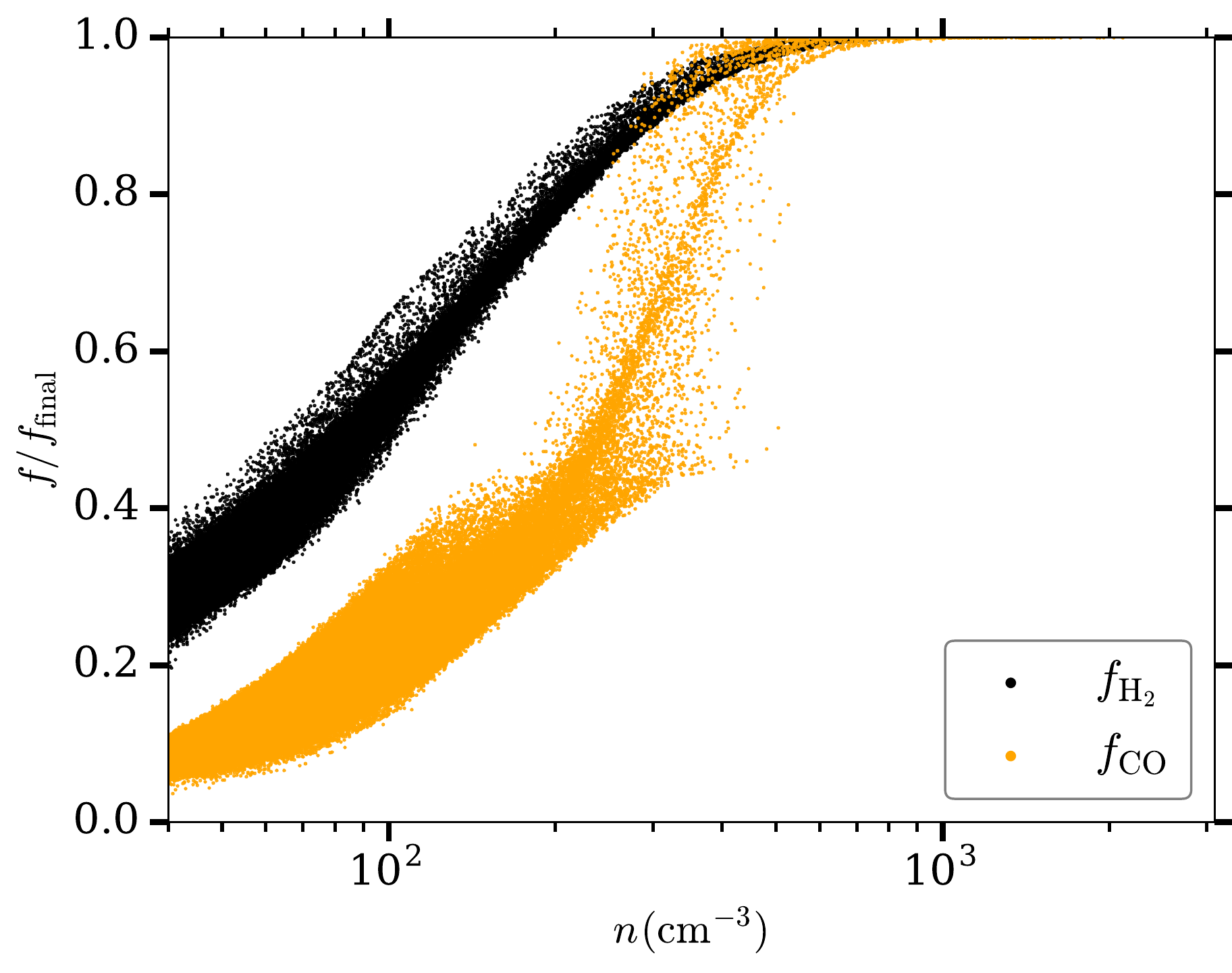}
\caption{Scatter plot of $f_\Ht$ (black) and $f_\CO$ (orange)
    versus density $n$ for model TCHEM-5Myr. 
    $f_\Ht$ and $f_\CO$ in each grid cell is normalized to the
    equilibrium abundance $f_\mr{H_2, final}$ and $f_\mr{CO,final}$ in that
    grid cell at $t_\mr{chem}=50~\Myr$ (in model TCHEM-50Myr).
}
\label{fig:H2_fCO_5Myr}
\end{figure}

The difference in the evolution of $\Ht$ and $\CO$ abundance 
comes from their different distributions.
As shown in Figure \ref{fig:H2_fCO_5Myr},
both $\Ht$ and $\CO$ abundances 
are closer to equilibrium at higher densities, as the rate of
collisional reactions increases with density.
In fact, at a given density in the
  range $\sim 40-400~\mr{cm^{-3}}$, at $5~\Myr$ the abundance of $\Ht$
  is closer than the abundance 
  of $\CO$ is to its final value.
However, in equilibrium most of the $\Ht$ is in gas at
intermediate densities
$n\approx 10-100~\mr{cm^{-3}}$, whereas most $\CO$ is in gas at high densities 
$n\gtrsim 200~\mr{cm^{-3}}$ (Figure \ref{fig:nH_hist_MH2_MCO}). This leads to
a shorter timescale for the overall $\CO$ abundance to reach equilibrium than
$\Ht$. Since the $\CO$ luminosity also increases much less than the $\Ht$ mass,
this leads to a lower
$X_\CO$ value at early $t_\mr{chem}$ (Table \ref{table:res_comparison}).

\begin{figure}[htbp]
\centering
\includegraphics[width=\linewidth]{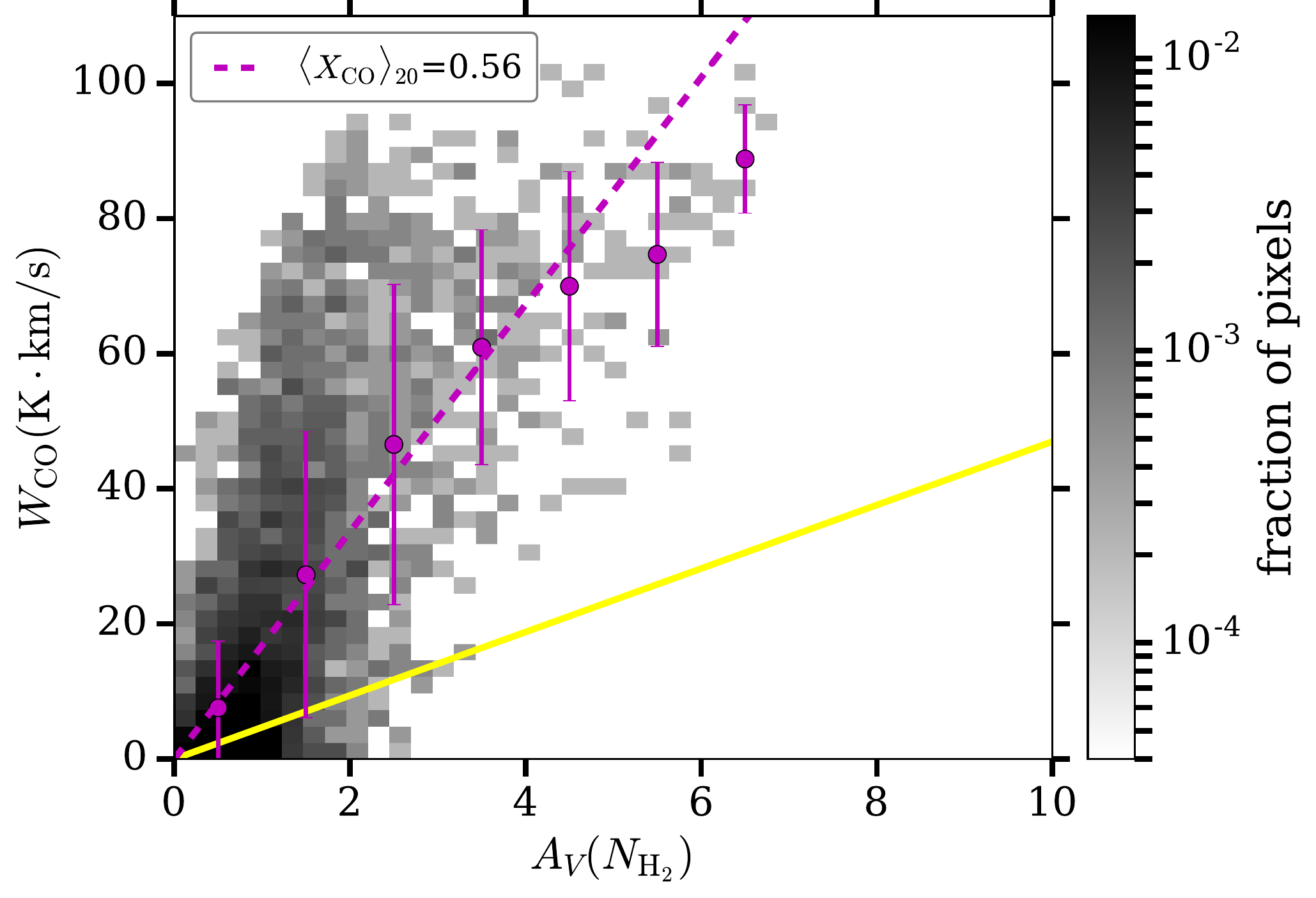}
\caption{Similar to the left panel of Figure \ref{fig:WCO_AVH2}, but for
model TCHEM-5Myr.}
\label{fig:WCO_AVH2_5Myr}
\end{figure}

Non-equilibrium chemistry also has an effect on the distribution of 
$W_\CO$ vs. $A_V(N_\Ht)$. For model TCHEM-5Myr (Figure \ref{fig:WCO_AVH2_5Myr}), 
the distribution of the pixels are shifted to the left compared to that in
TCHEM-50Myr (left panel of Figure \ref{fig:WCO_AVH2}). 
This is because $W_\CO$ is close to equilibrium, but $N_\Ht$ is a factor of $\sim 2$
smaller than the equilibrium values, for the same reasons discussed above.
Moreover, the distribution of $W_\CO$ vs. $A_V(N_\Ht)$ in TCHEM-5Myr shows some hints of
a plateau for $W_\CO$ at high $A_V(N_\Ht)$, especially in the binned average
value of $W_\CO$, which is not present in TCHEM-50Myr. This implies that
younger clouds may not only have lower $X_\CO$ on average, but also different
distributions of $W_\CO$ vs. $A_V(N_\Ht)$ compared to older clouds. We discuss
this further in Section \ref{section:result:2pc}.
Note that $N_\Ht$ includes all $\Ht$ along the line of sight, both in high density
clumps where $\CO$ forms, and in the foreground/background
low density envelopes with only $\Ht$ and no $\CO$.
Because most $\Ht$
(in equilibrium) 
lies in these low density envelopes, the fractions of
$\Ht$ in $\CO$-bright and $\CO$-dark regions increase by similar proportions
with $t_\mr{chem}$, and $f_\mr{dark}$ stays constant from 
$t_\mr{chem}=5~\Myr$ to $t_\mr{chem}=50~\Myr$ (Table \ref{table:res_comparison}).

\subsection{Variations in Galactic Environments\label{section:result:2pc}}
\begin{figure*}[htbp]
\centering
\includegraphics[width=0.9\linewidth]{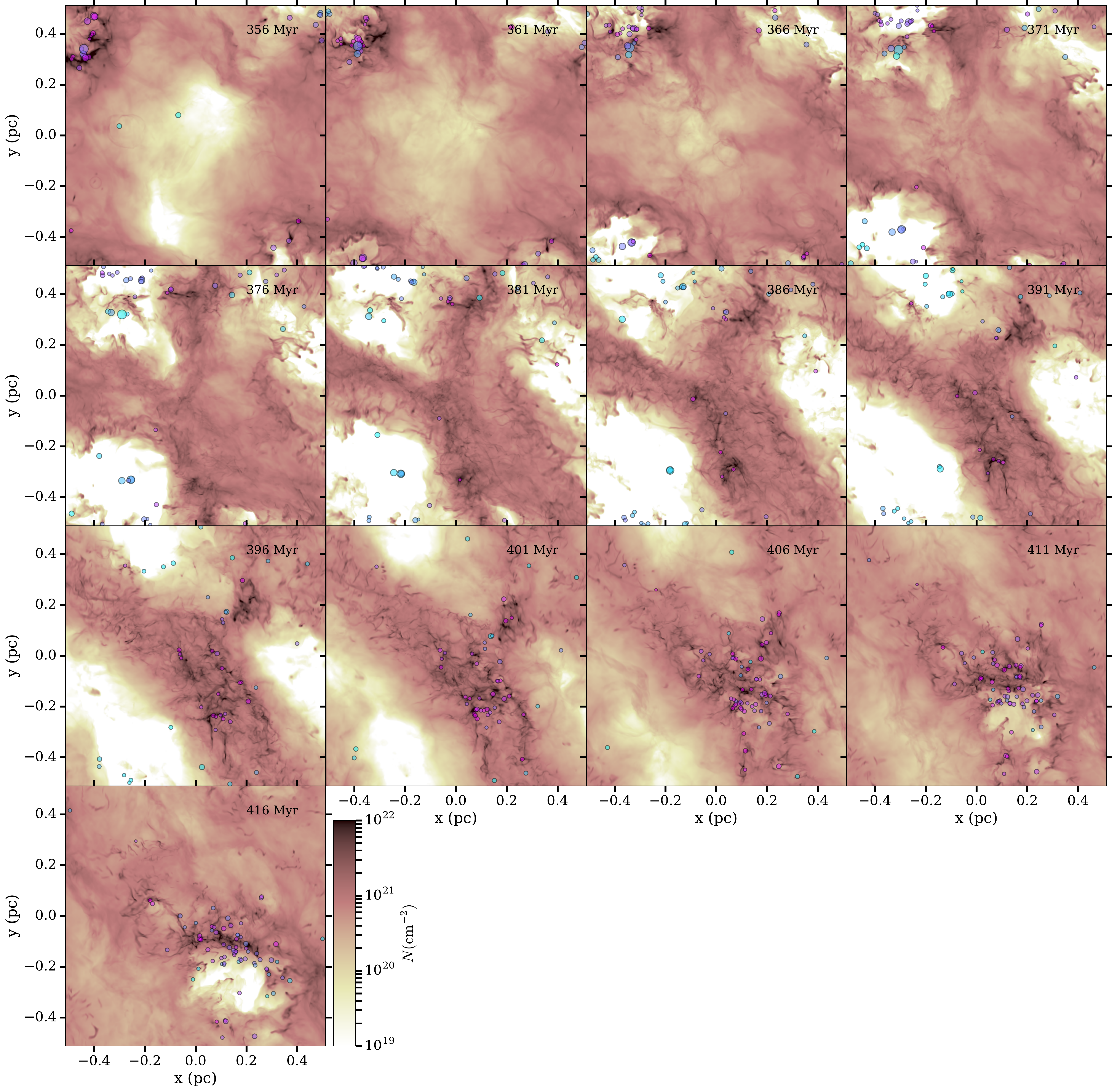}
\caption{Total gas surface density $N$ in models T-356Myr -- T-416Myr. The star
clusters are shown in circles, similar to Figure \ref{fig:res_overview} top
panels.}
\label{fig:NH_2pc}
\end{figure*}

Galactic environment fundamentally impacts the molecular content of the ISM.
Supernova feedback creates and destroys molecular clouds, shocks and turbulence
shape molecular clouds in different morphologies, and the
radiation field varies with the star formation activities. Some of these effects
can be seen visually in Figure \ref{fig:NH_2pc}. The morphology of molecular clouds
varies from dense concentrated structures (such as in T-356Myr), to more diffuse,
smaller clouds (such as in T-406Myr). The mass and number of young clusters also
changes over time, reflecting the variations in the star formation rate.
To quantify the effect of time-varying galactic environment on $X_\CO$, we compare 
models with $2~\pc$ resolution
at different times during the galactic evolution (model IDs start with T
in Table \ref{table:model}). As discussed in Section \ref{section:res_XCO}, 
the average $X_\CO$ is well resolved with a resolution of $2~\pc$ in
these simulations. 

A summary of models T-356Myr -- T-416Myr is listed in
Table \ref{table:2pc}. In these models, 
$M_\Ht$ and $L_\CO$ vary by factor of $\sim 3$, and the incident radiation field 
strength varies by a factor of $\sim 8$. However, 
despite these large variations in the environment, $\langle X_\CO \rangle$
stays almost constant, changing only by $\sim 40\%$. We found no strong correlation
(coefficient of determination $R^2<0.4$ in linear regression) between 
$\langle X_\CO \rangle$ and $M_\Ht$, the radiation field strength $\chi$, or 
the average extinction from $\Ht$ in $\CO$-bright
regions $\langle A_V \rangle_\mr{CO}$. \citet{Remy2017} measured
$\langle X_\CO \rangle$ in individual Milky Way molecular clouds 
using $\gamma$-ray observations, and they also found 
no strong correlation between $\langle X_\CO \rangle$ and $M_\Ht$ or 
$\langle A_V \rangle_\mr{CO}$.\footnote{\citet{Remy2017} shows a correlation between 
$\langle X_\CO \rangle$ and $\langle A_V \rangle_\mr{CO}$ with $R^2\approx 0.6$.
However, this relation is largely driven by one outlier, the Perseus cloud,
which has much lower $\langle X_\CO \rangle$ and higher $\langle A_V
\rangle_\mr{CO}$ than the rest of the sample. Excluding the Perseus cloud, we
found no strong correlation ($R^2\approx 0.3$) between 
$\langle X_\CO \rangle$ and $\langle A_V \rangle_\mr{CO}$ for the rest of their
sample.} 

\begin{figure*}[htbp]
\centering
\includegraphics[width=0.9\linewidth]{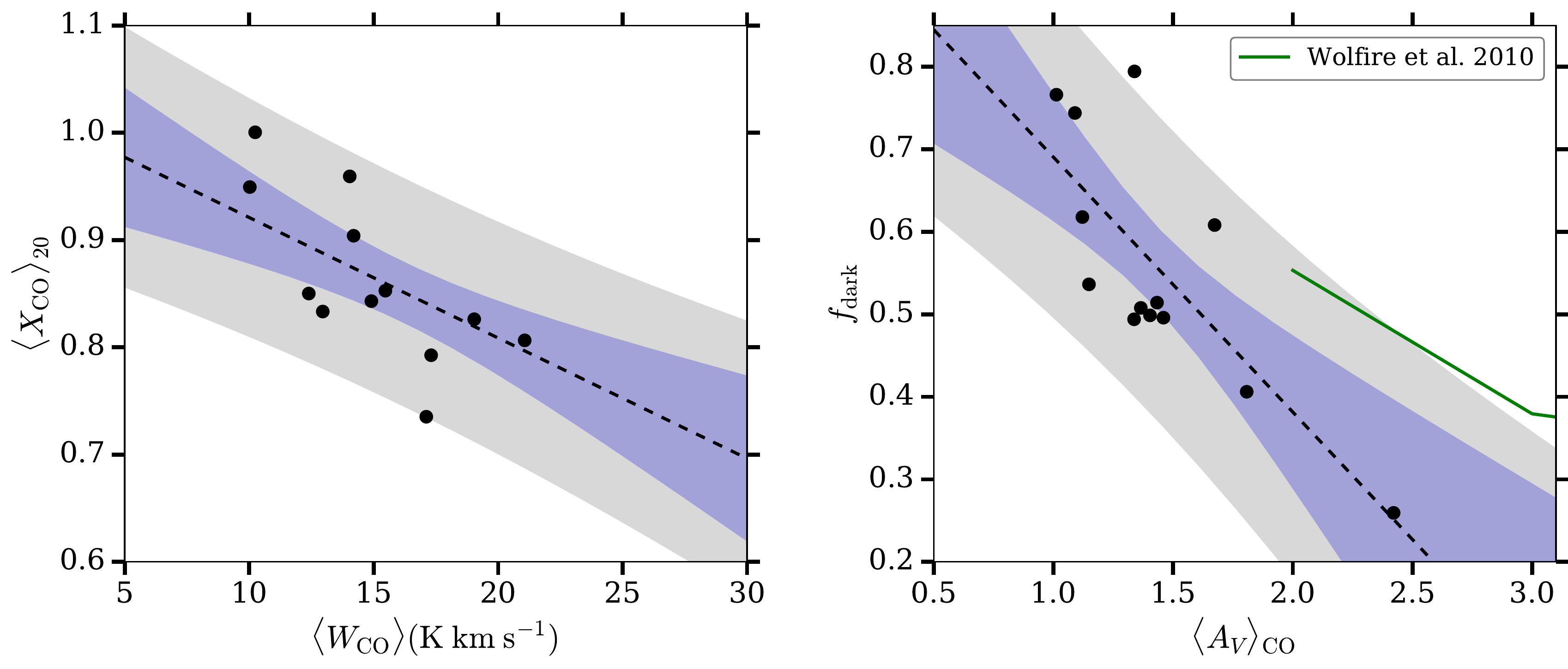}
\caption{$\langle X_\CO \rangle$ vs. $\langle W_\CO \rangle$ ({\sl left}), 
    and $f_\mr{dark}$ vs. $\langle A_V \rangle_\mr{CO}$ ({\sl right}) 
    in models T-356Myr -- T-416Myr.
    The dashed lines are the linear fits to the simulation data 
        (see text). The 90\% 
    confidence and prediction intervals from linear regression are indicated by
the purple and gray regions.
    In the right panel, the green line shows
    the theoretical model of spherical molecular clouds in \citet{WHM2010} (their
    Figure 11, with their definition of $\CO$-dark $\Ht$ translated to
        our definition of $f_\mr{dark}$ according to Equation (\ref{eq:fDG_fdark})).  }
\label{fig:summary_2pc}
\end{figure*}

  \citet{Remy2017} found a slight anti-correlation of
$\langle X_\CO \rangle$ and $\langle W_\CO \rangle$:
$\langle X_\CO \rangle \sim -0.051 \langle W_\CO \rangle$.
We similarly found a slight anti-correlation 
(Figure
\ref{fig:summary_2pc} left panel), with
$\langle X_\CO \rangle_{20} = -0.011 \pm 0.005 (\langle W_\CO
\rangle/\mr{K~km~s^{-1}}) + 1.0 \pm 0.09$, where the uncertainties represent the
90\% confidence intervals for the fitted slope and intercept. 
The slope of the linear fit is very shallow, 
and $\langle X_\CO \rangle$ is not sensitive to the
change of $\langle W_\CO \rangle$. 
We note however that \citet{Remy2017}
focuses on the nearby low mass molecular clouds with
much lower values of $\langle W_\CO \rangle \approx 2-10~\mr{K\cdot km/s}$
than $\langle W_\CO \rangle \approx 10-20~\mr{K\cdot km/s}$ 
in the GMCs in our simulations, and therefore may not be directly
comparable to our results.

Large scale galaxy simulations by 
\citet{Narayanan2012} found a similar trend that $\langle X_\CO \rangle$
decreases with increasing $\langle W_\CO \rangle$, although the range of
$\langle W_\CO \rangle$ is much larger in their simulations as they consider a
wide range of galactic environments. \citet{Narayanan2012} found that the
$\langle X_\CO \rangle$--$\langle W_\CO \rangle$ relation is caused by the
increase of gas temperature and velocity dispersion at high $\langle W_\CO
\rangle$, which leads to a faster increase of $W_\CO$ than $N_\Ht$, resulting
in the decrease of $X_\CO$. Similarly, we found that the snapshots in our
simulations with higher $\langle W_\CO \rangle$ 
also have larger velocity dispersions, 
although the gas temperature is roughly constant in
the $\CO$ forming regions in our models (see discussion of Figure 
\ref{fig:Texc_sigmav_WCO} in Section \ref{section:res_XCO}). Interestingly, this
is also consistent with the fact that the galactic center molecular clouds 
have larger velocity dispersions and lower $X_\CO$ compared to the solar
neighborhood clouds. We plan to carry out numerical simulations with
galactic-center-like environments 
in the future to study the variation of $X_\CO$ in
detail.

\begin{table*}[htbp]
    \caption{Overall properties of models: variations in galactic environment}
    \label{table:2pc}
    \begin{tabular}{c ccccc cccc}
        \tableline
        \tableline
        model ID &$M_\mr{tot}(M_\odot)$ &$M_\Ht(M_\odot)$ &$M_\CO(M_\odot)$
        &$L_\CO(\mr{K~km~s^{-1}pc^2})$
        &$\langle X_\CO \rangle_{20}$
        &$f_\mr{dark}$  &$f_{100}$ &$2\langle f_\Ht \rangle$
        &$\chi$\tablenotemark{a}
        \\ 
        \tableline
T356-Myr  &$8.02\times 10^6$ &$5.61\times 10^5$ &$4.52\times 10^2$  &$3.64\times 10^5$ &0.71   &26\%  & 4.5\%   &10\%  &3.0\\
T361-Myr  &$7.93\times 10^6$ &$4.25\times 10^5$ &$1.63\times 10^2$  &$1.96\times 10^5$ &0.81   &41\%  & 2.7\%   & 8\%  &1.8\\
T366-Myr  &$7.78\times 10^6$ &$3.38\times 10^5$ &$8.67\times 10^1$  &$1.00\times 10^5$ &0.83   &61\%  & 1.5\%   & 6\%  &1.1\\
T371-Myr  &$7.64\times 10^6$ &$3.03\times 10^5$ &$4.11\times 10^1$  &$5.31\times 10^4$ &0.74   &79\%  & 0.6\%   & 6\%  &0.9\\
T376-Myr  &$7.45\times 10^6$ &$5.34\times 10^5$ &$5.86\times 10^1$  &$8.23\times 10^4$ &0.95   &77\%  & 0.7\%   &10\%  &0.4\\
T381-Myr  &$7.41\times 10^6$ &$6.85\times 10^5$ &$8.19\times 10^1$  &$1.10\times 10^5$ &1.00   &74\%  & 0.9\%   &13\%  &0.4\\
T386-Myr  &$7.47\times 10^6$ &$8.54\times 10^6$ &$2.17\times 10^2$  &$2.40\times 10^5$ &0.85   &62\%  & 1.8\%   &16\%  &0.4\\
T391-Myr  &$7.59\times 10^6$ &$1.04\times 10^6$ &$3.77\times 10^2$  &$3.60\times 10^5$ &0.83   &54\%  & 2.6\%   &19\%  &0.4\\
T396-Myr  &$7.75\times 10^6$ &$9.25\times 10^6$ &$3.16\times 10^2$  &$3.47\times 10^5$ &0.84   &49\%  & 3.5\%   &17\%  &1.0\\
T401-Myr  &$7.97\times 10^6$ &$8.40\times 10^6$ &$2.73\times 10^2$  &$3.09\times 10^5$ &0.85   &50\%  & 3.6\%   &15\%  &1.4\\
T406-Myr  &$8.16\times 10^6$ &$6.82\times 10^5$ &$1.93\times 10^2$  &$2.16\times 10^5$ &0.96   &51\%  & 3.1\%   &12\%  &1.9\\
T411-Myr  &$8.29\times 10^6$ &$6.06\times 10^5$ &$1.68\times 10^2$  &$2.06\times 10^5$ &0.90   &51\%  & 2.4\%   &10\%  &1.4\\
T416-Myr  &$8.28\times 10^6$ &$5.51\times 10^5$ &$1.76\times 10^2$  &$2.19\times 10^5$ &0.79   &50\%  & 2.2\%   & 9\%  &1.0\\
        \tableline
average  &$7.83\times 10^6$  &$6.42\times 10^5$ &$2.00\times 10^2$ &$2.16\times 10^5$ &0.85  &56\% &2.3\% &12\% &1.2\\
        \tableline
        \tableline
    \end{tabular}
    \tablenotetext{1}{FUV radiation field intensity in \citet{Draine1978} units.}
\end{table*}

Unlike $\langle X_\CO \rangle$, the fraction of $\CO$-dark $\Ht$,
$f_\mr{dark}$, does show
significant variations and a strong correlation ($R^2=0.6$) with
$\langle A_V \rangle_\mr{CO}$ (Figure \ref{fig:summary_2pc} right panel). 
Linear regression gives $f_\mr{dark} = -0.31\pm 0.14 \langle A_V
\rangle_\mr{CO} + 1.0\pm 0.2$, where the uncertainties represent the
90\% confidence intervals for the fitted slope and intercept.
$f_\mr{dark}$ increases with decreasing $\langle A_V \rangle_\mr{CO}$. 
In other words, there is more $\CO$-dark $\Ht$ in more diffuse molecular
clouds, which is not surprising as $\CO$ forms in denser gas than $\Ht$.
The same trend was identified in the
simplified spherical molecular cloud model by \citet{WHM2010}.\footnote{The result
from \citet{WHM2010} shown in Figure \ref{fig:summary_2pc} 
is taken from their model with
metallicity $Z'=1.9$ and incident radiation field $\chi=10$. \citet{WHM2010}
found that $f_\mr{dark}$ is not sensitive to $Z'$ or $\chi$ in their studies.}
We note that \citet{WHM2010} uses a slightly different definition of $\CO$-dark
    $\Ht$, and we use Equation (\ref{eq:fDG_fdark}) to translate their
definition to ours.
We have also performed an experiment by running the
T-381Myr model only varying the radiation field strength, and found
$f_\mr{dark}$ stays constant over $\chi=0.4-3.5$, confirming the result from 
\citet{WHM2010} that $f_\mr{dark}$ is not sensitive to $\chi$.

\begin{figure}[htbp]
\centering
\includegraphics[width=0.9\linewidth]{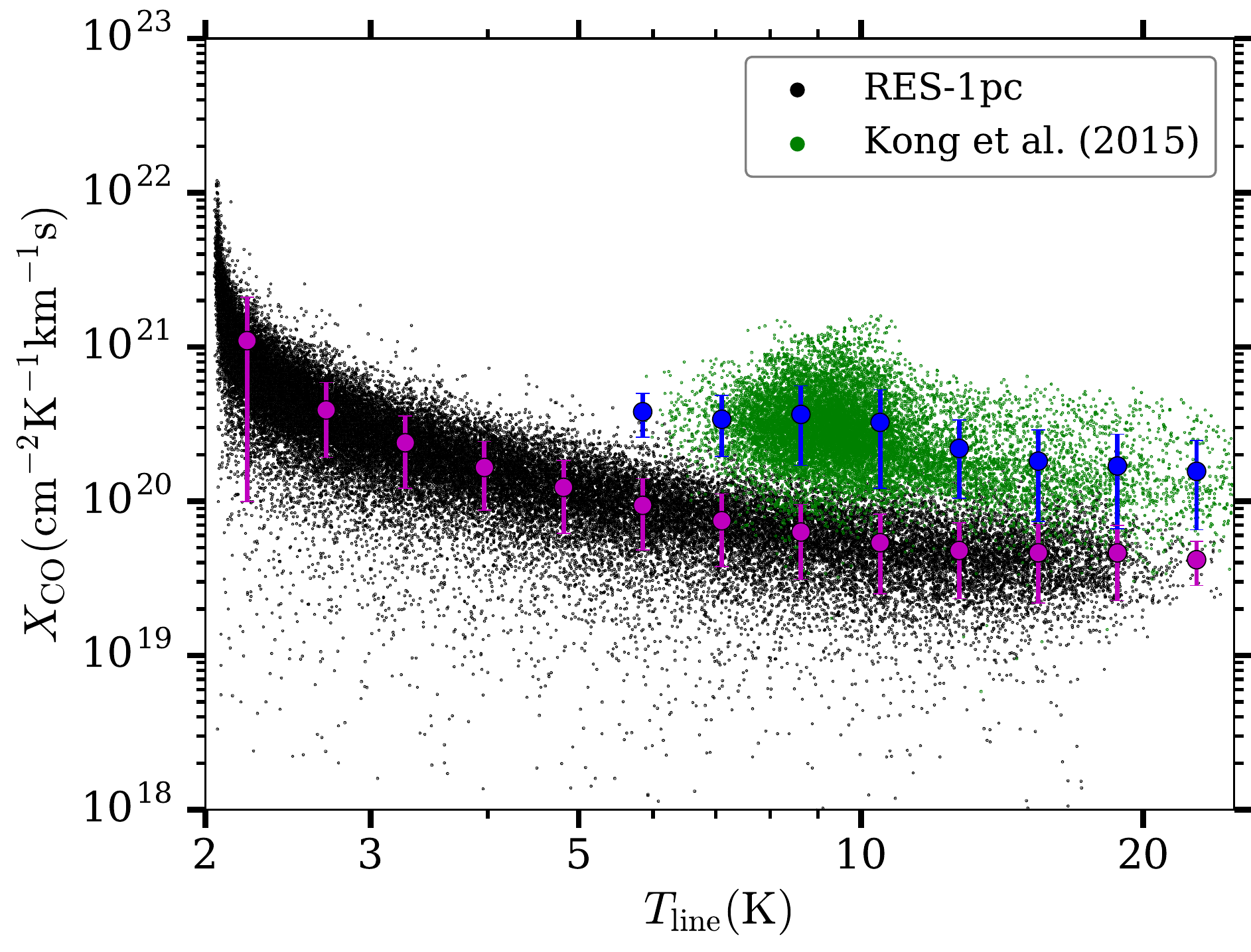}
\caption{Scatter plot of $X_\CO$ vs. $T_\mr{line}$ in models T-356Myr --
    T-416Myr (black) and \citet{Kong2015} (green). 
    Each point is one pixel in the simulations/observations.
The red and blue filled circles with error bars are the binned
mean values and standard deviations of $X_\CO$ in our simulations and
\citet{Kong2015}.}
\label{fig:XCO_Tline}
\end{figure}

Another comparison of $X_\CO$ with observations is shown in 
Figure \ref{fig:XCO_Tline}, where the $X_\CO$ in each pixel is plotted against
$T_\mr{line}$. Comparing to the California cloud observed by \citet{Kong2015},
our simulations shows a similar slope for the relation between $X_\CO$ and
$T_\mr{line}$ at $T_\mr{line}>6~\mr{K}$ (the observational data are not
available at lower $T_\mr{line}$). However, the 
value of $X_\CO$ at a given $T_\mr{line}$ is about a factor of
$\sim 4$ lower than the observations. One reason for this discrepancy may be
that \citet{Kong2015} observed $\CO(J=2-1)$ line and
assumed a fixed line ratio of $W_\CO(J=2-1)/W_\CO(J=1-0)=0.7$, and this ratio is
very uncertain. As discussed below in more detail, 
generally different observations and
also our simulations show similar trends for the variations in $X_\CO$, but the
absolute value of $X_\CO$ can differ by a factor of a few. 

\begin{figure}[htbp]
\centering
\includegraphics[width=\linewidth]{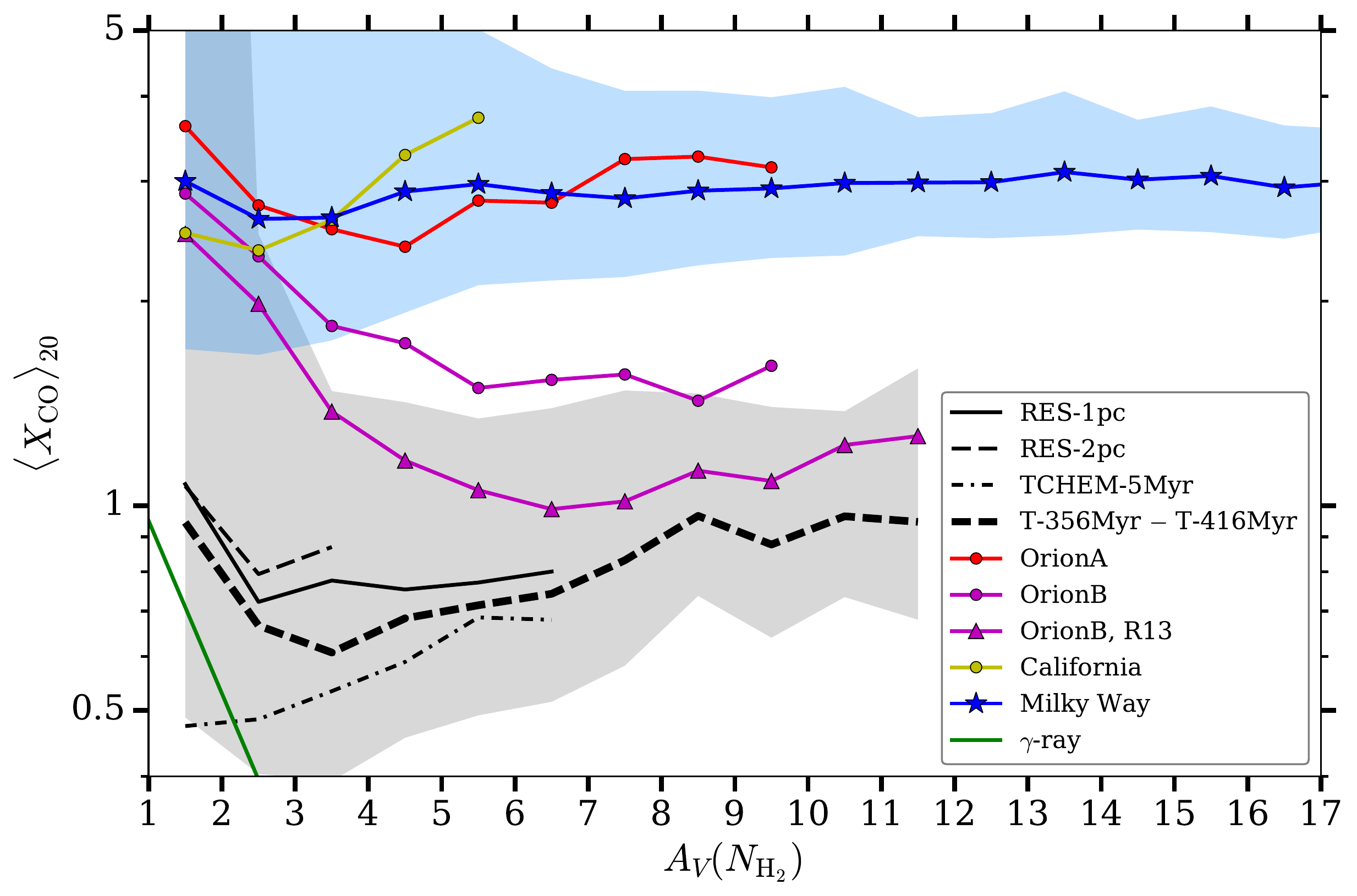}
\caption{The average $X_\CO$ binned in $A_V(N_\Ht)$. The black lines are the
    simulation models of RES-1pc (solid), RES-2pc (thin dashed), TCHEM-5Myr (dash
    dotted), and T-356Myr -- T-416Myr (thick dashed, 1-$\sigma$ dispersions
    showing as the gray shaded region). The filled circles show
    the observations of molecular clouds in \citet{Lee2018} for California
    (yellow), Orion A (red), and Orion B (magenta). The blue stars show the
    line-of-sight average of all clouds in \citet{Lee2018} (1-$\sigma$ dispersions
    showing as blue the shaded region). The magenta
    triangles show the observations of Orion A by \citet{Ripple2013}. The green
    solid line plots the $\gamma$-ray observations averaged over individual
    molecular clouds in \citet{Remy2017}.}
\label{fig:XCO_bin}
\end{figure}

Using all of the simulation models, a summary of $X_\CO$ as a function of
$A_V (N_\Ht)$ and comparison with observations is shown in Figure \ref{fig:XCO_bin}. 
Because of the large uncertainties in observations of $X_\CO$ 
at low $A_V (N_\Ht)$, we only plot the data at $A_V (N_\Ht) > 1$. 
Both in our simulations and the observations, there is a factor of $\sim 2$
variation in $X_\CO$ over $A_V (N_\Ht)=1-12$.
Simulations with $t_\mr{chem}=50~\Myr$ (RES-1pc,
RES-2pc, T-356Myr -- T-416Myr) show a
decrease of $X_\CO$ at $A_V (N_\Ht) \lesssim 3$, regardless of the resolution and
variations in galactic environments. Similar trends can be seen in the
observations of Orion molecular clouds by \citet{Lee2018} and \citet{Ripple2013}. 
In contrast, the TCHEM-5Myr model shows a flatter profile at
$A_V (N_\Ht) \lesssim 3$ and a slight increase of $X_\CO$ at $A_V (N_\Ht) > 3$.
Interestingly, the California cloud observed by \citet{Lee2018}
also shows a similar trend. Compared to Orion, the California cloud 
has similar mass and distance, but an order of magnitude lower star formation rate,
and therefore is believed to be much younger \citep{Lada2009}. 
This has interesting implications that the profile of $X_\CO$ as a function of
$A_V(N_\Ht)$ may be used as an indicator of the age of molecular clouds.

Although the trend for the correlation between $X_\CO$ and $A_V (N_\Ht)$ is
similar in our simulations and observations, there is a discrepancy in the
absolute value of $X_\CO$. This may be due to systematic
errors in either observations or simulations. One major uncertainty in
observations of $X_\CO$ comes from the assumptions in deriving $N_\Ht$.
Estimations of $\Ht$ based on $\gamma$-ray emission 
systematically give a factor of $~2$ lower $X_\CO$ than dust-based methods,
consistent with the value of $X_\CO$ in this paper 
\citep[][see also Figure \ref{fig:XCO_bin}]{BWL2013, Remy2017}.\footnote{The 
 observation by
\citet{Remy2017} in Figure \ref{fig:XCO_bin} is averaged over the molecular
clouds instead of individual pixels in a given $A_V(N_\Ht)$ range. Nonetheless, it
indicates the systematically lower $X_\CO$ in $\gamma$-ray observations.} Even
within the dust-based methods, the estimate of $X_\CO$ in Orion A 
based on dust emission 
is a factor of $\sim 2$ higher in observations of \citet{Lee2018} 
compared to that in \citet{Ripple2013} based on dust extinction.
As another example, the $X_\CO$ in Perseus measured 
by \citet{LeeMY2014} (dust emission) is a factor of $\sim 7$ lower than that
in \citet{Pineda2008} (dust extinction).

Several possible factors 
can contribute to the systematics in dust-based observations:
different assumptions of the dust to gas ratio, uncertainties in 
foreground/background subtraction, and different resolutions/beam size
(although the resolution effect is relatively mild, as
noted by \citet{LeeMY2014} and discussed in Section
\ref{section:result:beamsize}). \citet{LeeMY2014} discussed in detail for the
case of Perseus molecular cloud, that all these
factors can indeed lead to a different estimate of $X_\CO$. 
Sample differences in observations may also play a role. Most
observations of $X_\CO$ are for nearby low-mass star forming regions, while
most molecular clouds in the Milky Way and our simulations are forming or close
to high-mass stars. The feedback form high-mass stars may lead to slightly
higher velocity dispersions, and lower $X_\CO$. The only nearby high-mass star
forming molecular cloud is Orion, and it does have a lower value of
$X_\CO$ compared to the Milky Way average (Figures \ref{fig:WCO_AVH2} 
and \ref{fig:XCO_bin}).

For the numerical
simulations, the main uncertainties lie in the assumptions of equilibrium
chemistry and the sub-grid model of micro-turbulence in calculating the $\CO$
emission. 
As a further test,
  we produced synthetic observations of
model RES-2pc with half of the fiducial micro-turbulence velocity and no
sub-grid micro-turbulence (only thermal line-broadening on the grid scale), and
found that the values of $X_\CO$ increase by a factor of 1.4 and 1.8.
Therefore, the uncertainty in sub-grid micro-turbulence may account for part but not
all of the discrepencies in $X_\CO$ between our simulations and observations.
Future AMR simulations with higher numerical resolution and non-equilibrium chemistry
will be able to provide more insight into these issues. 

\subsection{Dependence of $X_\CO$ on the Observational Beam Size
\label{section:result:beamsize}}
Observation of molecular clouds often have different physical beam
sizes/resolutions, which depend on the telescope 
as well as the distance of the object.
In order to investigate the effect of observational resolution on $X_\CO$, we
smooth the synthetic observations to different beam sizes as described in
Section \ref{section:method:beamsize}.

\begin{figure}[htbp]
\centering
\includegraphics[width=\linewidth]{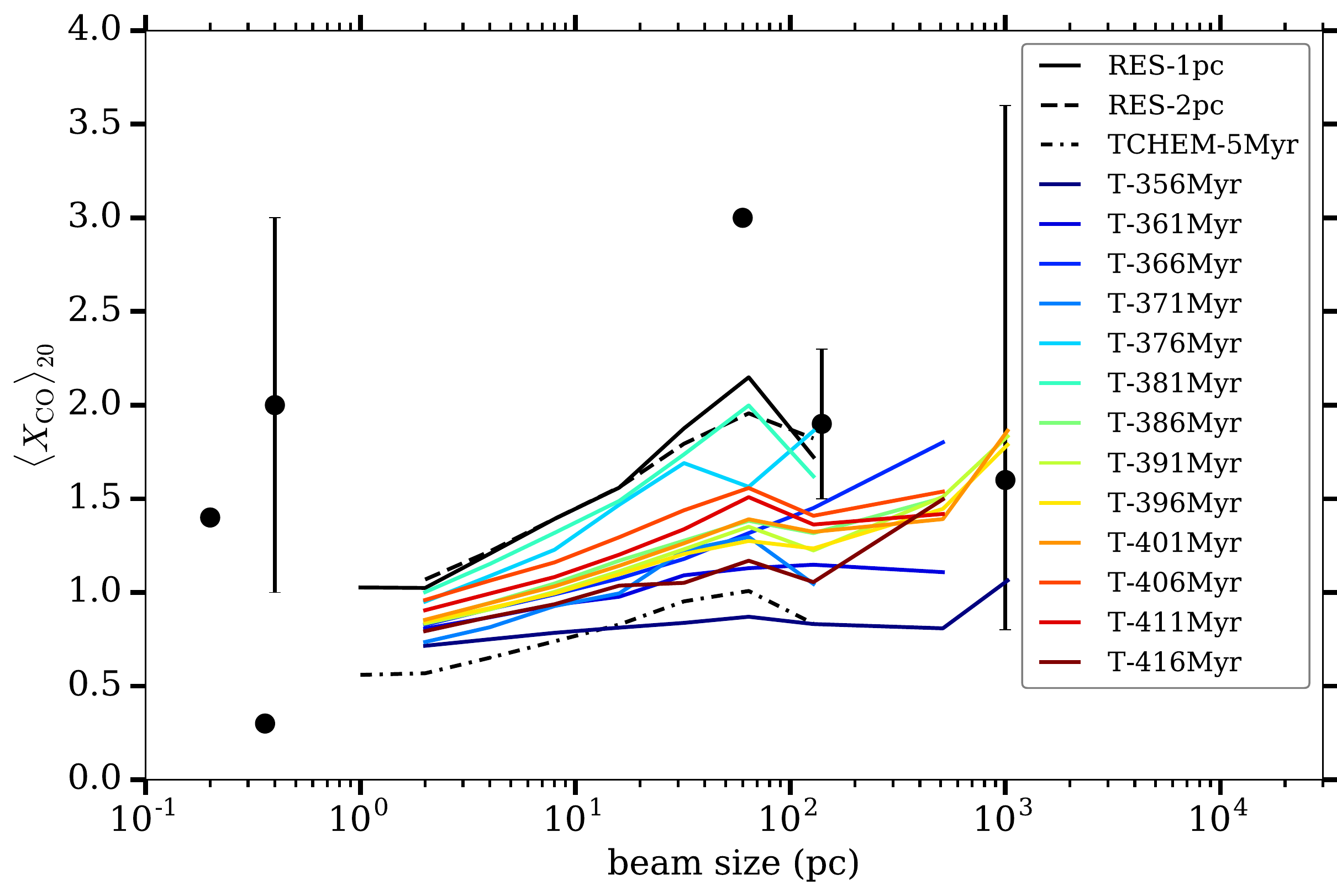}
\caption{$\langle X_\CO \rangle_{20}$ as a function of beam size in different
models (see label). The black circles (with error bars) are observations from
Table \ref{table:XCO_obs}.}
\label{fig:XCO_pixsize}
\end{figure}

\begin{figure}[htbp]
\centering
\includegraphics[width=0.9\linewidth]{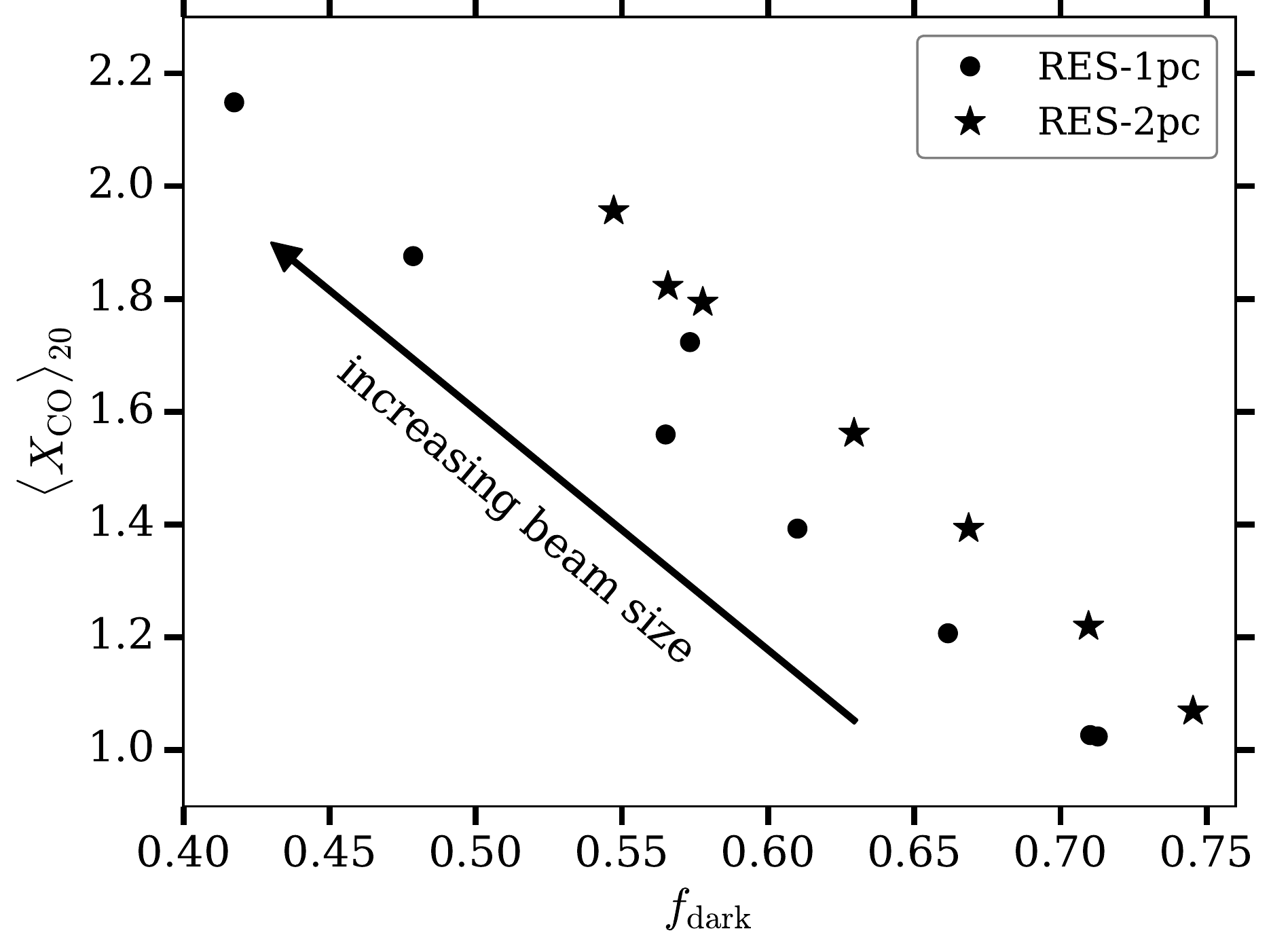}
\caption{$\langle X_\CO \rangle_{20}$ vs. $f_\mr{dark}$ in models RES-1pc and
RES-2pc. Each point is for a different beam size in Figure \ref{fig:XCO_pixsize}.}
\label{fig:XCO_fdark_res}
\end{figure}

$\langle X_\CO \rangle$ increases by a factor of $\sim 2$ as the beam size
increases from $\sim 1~\pc$ to $\sim 100~\mr{pc}$, 
as shown in Figure \ref{fig:XCO_pixsize}.
This is a result of
the $\CO$-dark $\Ht$. The total $\CO$ emission remains the same as the beam
size increases, because the detection limits for different beam sizes (Table
\ref{table:para_obs}) are generally sensitive enough to detect most of the $\CO$
emission. This is not surprising as the sensitivity in observations are
designed to serve the purpose of accurately measuring the $\CO$ emission.
However, the $\CO$ emission is smoothed out spatially as the beam size
increases, resulting in a larger area of $\CO$-bright regions. Although the total
mass of $\Ht$ remains the same, because $X_\CO$ is calculated only within
$\CO$-bright regions, a larger area of $\CO$-bright regions leads to a larger
fraction of $\Ht$ mass accounted for, and therefore an increase of $X_\CO$. 
This is clearly illustrated in Figure \ref{fig:XCO_fdark_res}, showing the
correlation between $f_\mr{dark}$ and $\langle X_\CO \rangle$. 
From beam sizes of $\sim 100~\pc$ to $\sim 1~\mr{kpc}$, some simulations show
a continued increase of $X_\CO$ (e.g. T-401Myr), but some simulations with
more diffuse molecular clouds (e.g. T-381Myr) start to have part or all of
their
$\CO$ emission falling below the detection limits, leading to a non-detection
of $W_\CO$ or reduction of $X_\CO$. This suggests that some diffuse molecular
clouds may not be detected with a beam size coarser than $\sim 100 ~\pc$ in
extragalactic observations.

In Figure \ref{fig:XCO_pixsize}, 
we plot the observations of $\langle X_\CO \rangle$ in 
Milky Way molecular clouds and nearby galaxies (Table \ref{table:XCO_obs}). 
Because of the large uncertainties in the observations (as discussed above, and 
also seen directly in the different $\langle X_\CO \rangle$
from two Perseus observations) and dispersions
of $\langle X_\CO \rangle$ in different molecular clouds,
we cannot identify any obvious trend for the $X_\CO$ variation with beam size.
The general range of $X_\CO$ in the simulations is similar to the
observations.

\section{Summary}\label{section:summary}
In this paper, we theoretically model the $X_\CO$ conversion
factor by post-processing MHD galactic disk ISM simulations with chemistry and
radiation transfer to produce synthetic observations of molecular clouds.
We conduct detailed analyses of the dependence of molecular abundances and
observed line strengths on ISM conditions, and also consider numerical and
observational effects on calculated and measured $X_\CO$.
Our main findings are as follows:

\begin{enumerate}
    \item $\CO$ is only a very approximate tracer of $\Ht$. In our
        simulations, most $\Ht$ forms at 
        intermediate densities $n\approx 10-100~\mr{cm^{-3}}$, 
        but most $\CO$ forms 
        at higher densities $n\gtrsim 200~\mr{cm^{-3}}$ 
        (Figure \ref{fig:nH_hist_MH2_MCO}). The $\Ht$ abundance is
        determined mostly by density, while the $\CO$ abundance by dust
        shielding (Figures \ref{fig:fH2_n_AVsh}, \ref{fig:fCO_n_AVsh}).
        With a $2~\pc$ numerical resolution, $\Ht$ abundance is
        converged, but $\CO$ is not. Although there is considerable scatter, 
        the mean relation between the $\CO$ and $\Ht$
        column densities in the simulations are in agreement with 
        observations of UV absorption spectra (Figure \ref{fig:NCO_AV}).
    \item For $\CO$ emission, the high optical depth of the line further
        complicates the observable relation to $\Ht$. 
        On parsec scales, $W_\CO$ is largely determined by the mean
        excitation temperature of $\CO$ (Figure \ref{fig:Texc_sigmav_WCO}),
        which is in turn determined by the mean gas density. 
        Thus, $W_\CO$ most directly probes the mean gas density along the line
        of sight. However, for the turbulent clouds in our simulations, the
        mass-weighted mean volume density along a line of sight tends to be
        correlated with column density. This leads to a correlation between 
        $W_\CO$ and $N_\Ht$ (Figure \ref{fig:WCO_nHM_AVH2_hist}).
    \item A numerical resolution of at least $2~\pc$ is needed in order to
        resolve the average $X_\CO$ in molecular clouds for solar neighborhood
        conditions (Figure \ref{fig:XCO_hist}). In our simulations with
        environmental conditions similar to the solar neighborhood, we
        found $\langle X_\CO \rangle = 0.7-1.0 \times 10^{20}~
        \mr{cm^{-2}K^{-1}km^{-1}s}$, about a factor of 2 lower than the
        estimate from dust-based observations, and consistent with the
        $X_\CO$ from $\gamma$-ray observations. The value of 
        $\langle X_\CO \rangle$ is not sensitive to the 
        variations in molecular cloud mass,
        extinction, or the strength of the FUV radiation field 
        (Table \ref{table:2pc}).
    \item We found the $\CO$-dark $\Ht$ fraction $f_\mr{dark}=26-79\%$, which
        has an anti-correlation with the average extinction of molecular
        clouds (Figure \ref{fig:summary_2pc} right panel).
    \item The chemical timescale for $\Ht$ abundance to reach 
        equilibrium is longer than that for $\CO$ (Figure
        \ref{fig:fCO_fH2_all_t}), because of differences in characteristic
        densities. As a result, younger
        molecular clouds are expected to have lower $\langle X_\CO \rangle$
        values and flatter profiles of $X_\CO$ versus extinction
        compared to older molecular clouds 
        (Figures \ref{fig:WCO_AVH2_5Myr}, \ref{fig:XCO_bin}).
    \item As the observational beam size increases from $\sim 1~\pc$ to
        $\sim 100~\pc$, $\langle X_\CO \rangle$ increases by a factor of 
        $\sim 2$, due to the decrease of the $\CO$-dark $\Ht$ fraction (Figures
        \ref{fig:XCO_pixsize}, \ref{fig:XCO_fdark_res}).
    \item Our numerical simulations successfully reproduce the observed 
        variations of $W_\CO$ on parsec scales, as well as the 
        trends for the dependence of
        $X_\CO$ on extinction and the $\CO$ excitation temperature.
        However, the value of $X_\CO$ in our simulations is
        systematically lower by a factor of $\sim 2$ compared to dust-based
        observations (Figures 
    \ref{fig:WCO_AVH2}, \ref{fig:XCO_Tline}, \ref{fig:XCO_bin}). 
\end{enumerate}

The overall
agreement between our numerical simulations and observations of Milky Way
molecular clouds give us
confidence that similar simulations can be used to probe the $X_\CO$
conversion factors in different environments, such as the Galactic center, 
low metallicity dwarfs, and extreme star-forming systems (ultra luminous
infrared galaxies and high redshift galaxies). 
In a follow-up study, we will investigate the
properties of individual molecular clouds in our simulations.
In the future, we also plan to integrate full non-equilibrium
chemistry with the MHD simulations. 

\section{Acknowledgment}
This work was supported by grants NNX14AB49G from NASA, and AST-1312006
and AST-1713949 
from the NSF.
We thank the referee for helping us to improve
  the overall quality of this paper,
Jim  Stone, Kengo Tomida and Christopher J. White for
making the code {\sl Athena++} available and their help in developing the chemistry
module in the code, Mark H. Heyer for
providing the observational data from \citet{Ripple2013}, Adam K. Leroy for
providing the data from \citet{Lee2018} and helpful
discussions of comparisons with observations, Shuo Kong for providing the data
from \citet{Kong2015}, and Simon C. O. Glover for
suggesting an investigation in the effects of the numerical resolution.

\appendix
\section{Definitions of $\CO$-dark $\Ht$}
In this paper, we define the $\CO$-dark $\Ht$ as the molecular gas without
$\CO$ emission {\sl along a given line of sight}. \citet{WHM2010} uses a slightly
different definition in their spherical cloud model, 
and they refer the $\CO$-dark $\Ht$ as the molecular gas
outside of the optical depth $\tau_\mr{CO}=1$ surface. 
Their definition of $\CO$-dark $\Ht$ includes the $\Ht$
along the line of sight in the projected $\CO$-bright areas on the plane of the
sky, as long as it is outside
of the $\tau_\mr{CO}=1$ surface (see their Figure 1). In other words, the definition of
\citet{WHM2010} is in 3D physical space while our definition is in 2D
observational space.

To compare the result from \citet{WHM2010} to our simulations, we need to
translate their definition of $\CO$-dark $\Ht$ fraction, denoted by $f_\mr{DG}$
(their Equation 1) to our definition denoted by $f_\mr{dark}$ (Equation
(\ref{eq:f_dark}) in this paper). Below we derive the relation between
$f_\mr{DG}$ and $f_\mr{dark}$. We refer the readers to Figure 1 in
\citet{WHM2010} for a useful illustration for this derivation.

From Equation (\ref{eq:f_dark}), $f_\mr{dark}$ can be written as:
\begin{equation}\label{eq:f_dark_Mbr}
    f_\mr{dark} = \frac{M_\Ht-M_\mr{br}}{M_\Ht} = 1 -
    \frac{M_\mr{br}}{M_\Ht},
\end{equation}
where $M_\Ht$ is the total $\Ht$ mass (same as $M_\mr{H_2, tot}$ 
in Equation (\ref{eq:f_dark})), $M_\mr{br}$ is the mass in $\CO$-bright
areas on the projected sky. From Figure 1 in \citet{WHM2010}, 
$M_\mr{br}=M_\mr{CO} + M_\mr{DG}$, where $M_\CO$ is the mass with $r < R_\CO$,
and $R_\CO$ is the radius of the cloud where $\tau_\CO = 1$.
$M_\mr{DG}$ is the mass that lies within $R_\CO$ in the 2D projected sky,
but outside $R_\CO$ in the 3D cloud. Compared to the
definition in \citet{WHM2010},
\begin{equation}\label{eq:f_DG}
f_\mr{DG} = 1- \frac{M_\CO}{M_\Ht},
\end{equation}
$M_\mr{DG}$ is the part of the cloud that \citet{WHM2010} considered to be
$\CO$-dark, but we do not.

\citet{WHM2010} assumes
the cloud has a density profile $n(r) = n_0 (r_0/r)$, where $n_0$ and $r_0$ are
constants. This gives:
\begin{equation}\label{eq:M_H2}
    M_\Ht = \int_0^{R_\Ht} 4\pi m_\mr{H} n r^2 \di r = 2\pi n_0 m_\mr{H} r_0
    R_\Ht^2,
\end{equation}
and similarly,
\begin{equation}\label{eq:M_CO}
    M_\CO = 2\pi n_0 m_\mr{H} r_0 R_\CO^2,
\end{equation}
where $m_\mr{H}$ is the mass of the hydrogen atom.
$M_\mr{DG}$ can be estimated by
$M_\mr{DG} \approx 2\pi R_\CO^2 \Sigma_\mr{DG}$,
where $\Sigma_\mr{DG} = m_\mr{H} \int_{R_\CO}^{R_\Ht} n \di r = n_0 m_\mr{H}
r_0 \ln(R_\Ht/R_\CO)$. Therefore,
\begin{equation}\label{eq:M_br}
    M_\mr{br} = M_\CO + M_\mr{DG} \approx 2\pi n_0 m_\mr{H} r_0 R_\CO^2\left[1 +
    \ln\left( \frac{R_\Ht}{R_\CO} \right) \right].
\end{equation}
Equations (\ref{eq:f_dark_Mbr}) -- (\ref{eq:M_br}) then gives the relation
between $f_\mr{DG}$ and $f_\mr{dark}$:
\begin{equation}\label{eq:fDG_fdark}
    f_\mr{dark} \approx f_\mr{DG} - \frac{1}{2}(1-f_\mr{DG})\ln
    \left(\frac{1}{1-f_\mr{DG}} \right).
\end{equation}

\section{Test of the RADMC-3D code}
\begin{figure}[htbp]
\centering
\includegraphics[width=0.5\linewidth]{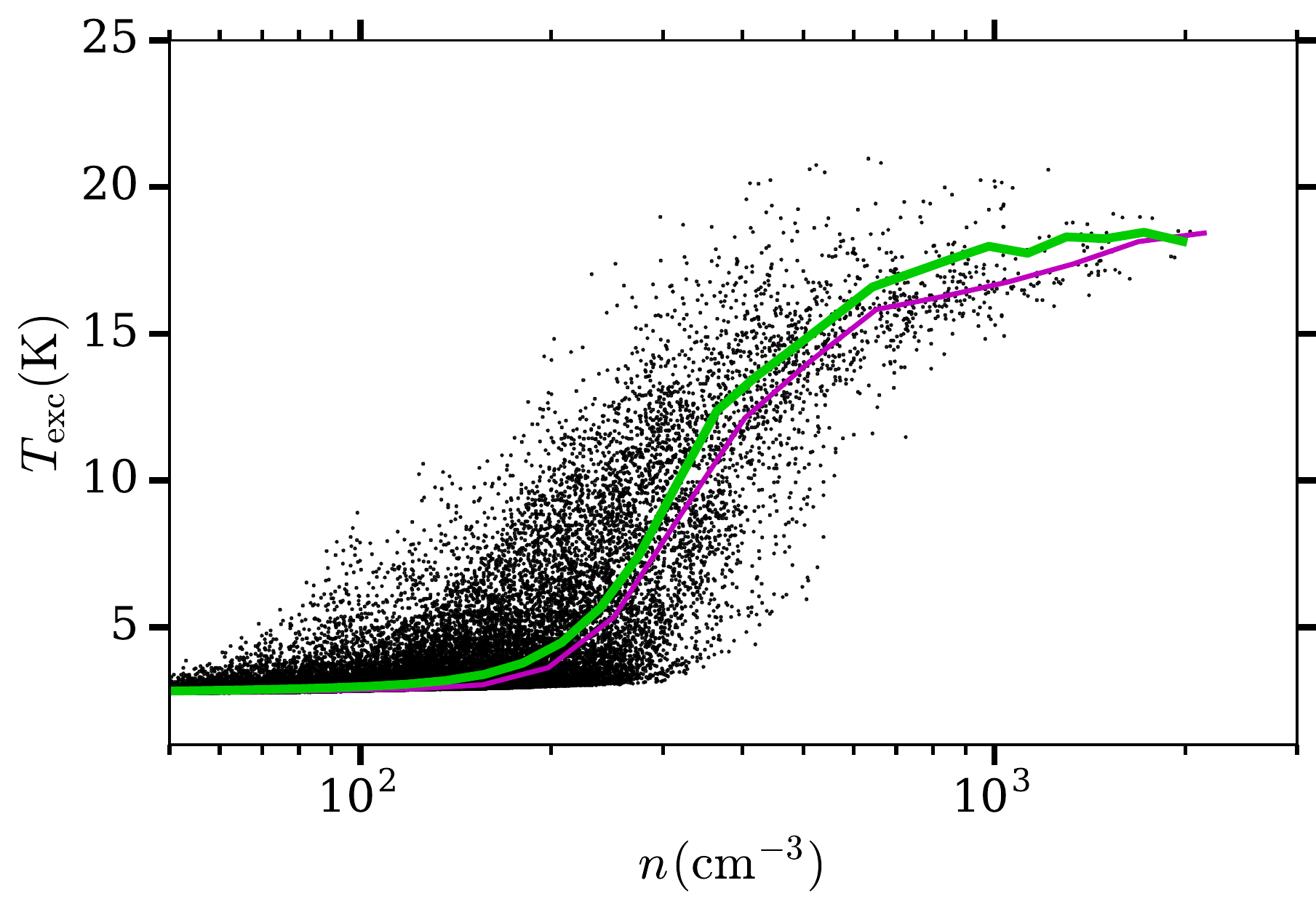}
\caption{Comparison of the excitation temperature
from the RADMC-3D radiation transfer code (scatter points with median values in
density bins shown as the magenta line) to the analytical 2-level system model (green
line) for the simulation RES-1pc. Only the $J=0$ and $J=1$ rotational levels of $\CO$ are
included in the calculations using RADMC-3D.}
\label{fig:Texc_nH_2level}
\end{figure}

Figure \ref{fig:Texc_nH_2level} shows a test for the RADMC-3D radiation
transfer code.  The level populations of $\CO$ are solved with only the first
two rotational levels instead of the default 41 levels. The analytical model
uses Equations (\ref{eq:T_exc}) and (\ref{eq:n12n0}) to compute $T_\mr{exc}$
versus $n$. Note that because $\tau_\mr{LVG}$ depends on level populations (see
Equation (\ref{eq:tau_LVG})), the average values of $\tau_\mr{LVG}$ in this case 
are slightly larger than that given by Equation (\ref{eq:tau_LVG_fit}). 

\bibliographystyle{apj}
\bibliography{apj-jour,thesis}
\end{document}